\documentclass{aa}  

\usepackage{graphicx}
\usepackage{txfonts}
 \usepackage{subcaption}         
 \usepackage{lscape}             
 \usepackage{placeins}           
                                
\usepackage{hyperref}
\hypersetup{
    colorlinks=true,
    citecolor=blue,
    linkcolor=blue,
    urlcolor=blue
    }
\usepackage{siunitx}
\usepackage{booktabs}

\newcommand{\hi}{\textrm{H\textsc{i}}}

\usepackage{color}
\definecolor{darkgreen}{rgb}{0.0, 0.5, 0.0}

\usepackage{stfloats} 

\begin{document}

   \title{Hydrogen intensity mapping with MeerKAT: Preserving cosmological signal by optimising contaminant separation\thanks{On behalf of the MeerKLASS Collaboration}}

    \titlerunning{Contaminant separation for MeerKAT \hi~intensity mapping}


   \author{
        I. P. Carucci\inst{1,2}
        \fnmsep\thanks{E-mail: \href{mailto:isabella.carucci@inaf.it}{isabella.carucci@inaf.it}}
          \and
          J. L. Bernal\inst{3}
          \and
          S. Cunnington\inst{4}
          \and
          M. G. Santos\inst{5}
          \and
          J. Wang\inst{6,5}
          \and
          J. Fonseca\inst{7,5}
          \and
          K. Grainge\inst{4}
          \and
          M. O. Irfan\inst{8}
          \and
          Y. Li\inst{9,5}
          \and
          A. Pourtsidou\inst{10,11}
          \and
          M. Spinelli\inst{12,5}
          \and
          L. Wolz\inst{4}
          }
    
    \institute{INAF - Osservatorio Astronomico di Trieste, Via G.B. Tiepolo 11, 34131 Trieste, Italy
    \and IFPU - Institute for Fundamental Physics of the Universe, Via Beirut 2, 34151 Trieste, Italy
    \and Instituto de Física de Cantabria (IFCA), CSIC-Univ. de Cantabria, Avda. de los Castros s/n, E-39005 Santander, Spain
    \and Jodrell Bank Centre for Astrophysics, Department of Physics \& Astronomy, The University of Manchester, Manchester M13 9PL, UK 
    \and Department of Physics and Astronomy, University of the Western Cape, Robert Sobukwe Road, Cape Town 7535, South Africa 
    \and Shanghai Astronomical Observatory, Chinese Academy of Sciences, 80 Nandan Road, Shanghai, 200030, China 
    \and Instituto de Astrofisica e Ci\^{e}ncias do Espa\c{c}o, Universidade
do Porto CAUP, Rua das Estrelas, PT4150-762 Porto, Portugal 
    \and Institute of Astronomy, Madingley Road, Cambridge CB3 0HA, UK 
    \and Liaoning Key Laboratory of Cosmology and Astrophysics, College of Sciences, Northeastern University, Shenyang 110819, China 
    \and Institute for Astronomy, University of Edinburgh, Royal Observatory, Blackford Hill, Edinburgh, EH9 3HJ, UK 
    \and Higgs Centre for Theoretical Physics, School of Physics and Astronomy, University of Edinburgh, Edinburgh EH9 3FD, UK 
    \and Observatoire de la Côte d’Azur, Laboratoire Lagrange, Bd de l’Observatoire, CS 34229, 06304 Nice cedex 4, France }

\authorrunning{I. P. Carucci et al.}
   \date{Received December XX, 2024}

\abstract{Removing contaminants is a delicate, yet crucial step in neutral hydrogen (\hi) intensity mapping and often considered the technique's greatest challenge. 
Here, we address this challenge by analysing \hi~intensity maps of about $100$ deg$^2$ at redshift $z\approx0.4$ collected by the MeerKAT radio telescope, an SKA Observatory (SKAO) precursor, with a combined  10.5-hour observation. Using unsupervised statistical methods, we removed the contaminating foreground emission and systematically tested, step-by-step, some common pre-processing choices to facilitate the cleaning process.
We also introduced and tested a novel multiscale approach: the data were redundantly decomposed into subsets referring to different spatial scales (large and small), where the cleaning procedure was performed independently. We confirm the detection of the \hi~cosmological signal in cross-correlation with an ancillary galactic data set, without the need to correct for signal loss. In the best set-up we achieved, we were able to constrain the \hi~distribution through the combination of its cosmic abundance ($\Omega_{\hi}$) and linear clustering bias ($b_{\hi}$) up to a cross-correlation coefficient ($r$). We measured $\Omega_{\hi}b_{\hi}r = [0.93 \pm 0.17]\,\times\,10^{-3}$ with a $\approx6\sigma$ confidence, which is independent of scale cuts at both edges of the probed scale range ($0.04 \lesssim k \lesssim 0.3 \,h$ Mpc$^{-1}$), corroborating its robustness.
Our new pipeline has successfully found an optimal compromise in separating contaminants without incurring a catastrophic signal loss. This development instills an added degree of confidence in the outstanding science we can deliver with MeerKAT on the path towards \hi~intensity mapping surveys with the full SKAO.}


   \maketitle

\section{Introduction}

Remarkably promising yet challenging to detect, mapping the cosmos with the fluctuations in 21-cm radiation from neutral hydrogen (\hi) is set to revolutionise the study of the Universe's large-scale structure \citep{Bharadwaj2001,Chang2008,Loeb2008}. The excellent redshift resolution of  intensity maps covering vast areas enables us to characterise the structures' expansion history and growth, providing constraints on its dark energy, dark matter, and neutrino mass properties \citep{Bull2015,Villaescusa-Navarro2017,Carucci2018,Obuljen2018,Witzemann2018,Berti2022}.

However, at the same frequencies of the redshifted 21-cm line, other astrophysical sources, such as our Galaxy, shine with considerably higher intensities (by four to five orders of magnitude), adding gigantic foregrounds to the sought-after signal \citep{Ansari2012,alonso2014}. That huge dynamic range between the cosmological signal and the foregrounds makes any tiny miscalibration or instrumental systematics produce catastrophic leakages, mixing up signals, and rendering their separation rather difficult \citep{Shaw2015,carucci2020,Matshawule:2020fjz,Wang2022}.

Nevertheless, many instruments and collaborations worldwide are actively involved in (or have already completed) \hi~intensity mapping surveys. Some of them have successfully detected the signal by using secondary tracer information \citep{Pen2009,chang2010,masui2013,switzer2013,wolz2017,Anderson2018,Tramonte2020,Li2021,wolz2022,CL23,Chime2023,Chime2024,MK_GAMA}, keeping pace with more and more stringent upper limits \citep{ghosh2011,Chakraborty2021,Pal2022,uGMRT2024}, and even detecting the signal on its own, albeit at small (non-cosmological) scales \citep{Paul2023}. 
Finally, other radio telescopes have been (sometimes purposely) planned to perform radio intensity mapping and either construction or systematic characterisation is ongoing \citep{tianlai,Hu2020,hirax,bingo,fast}.

In the observational works cited above, the instruments used are varied (e.g. single-dish radio telescopes, phased array feeds, packed interferometers, and arrays used as a collection of single dishes). 
Contaminant removal strategies, too, differ substantially. Some teams have opted for foreground `avoidance' and tried to filter the subset of the data with the strongest cosmological signal. Others use what the signal-processing community refers to as  `blind source separation' algorithms to try to be as agnostic as possible to the nature of the contaminants. Here, we opt for the latter.

This article focuses on the contaminant removal strategy adopted by the MeerKLASS collaboration, which runs \hi~intensity map observational campaigns with the MeerKAT telescope \citep{Santos2016,wang2021,MK_GAMA}. In particular, our starting point is the detection of signal in cross-correlation between MeerKAT radio intensity maps and galaxies from the WiggleZ Dark Energy Survey at redshift $z\approx0.4$ reported in \citet{CL23}, which we  refer to as CL23 in the remainder of the text.

In theory, astrophysical foregrounds are smooth along frequency, compared to the noisy spectral structure of the \hi~signal; hence, they are easily separable from the signal, for instance, via a principal component analysis (PCA), holding limited spectral degrees of freedom. In practice, bandpass fluctuations, chromatic beam response, and leakage of polarised foregrounds into the unpolarised signal  render the spectrally smooth assumptions invalid. 
In this work, we address whether the hypotheses  
behind PCA-like methods are still satisfied with observations and how to pre-process the data to make contaminant removal more efficient.

Building up an efficient cleaning pipeline is challenging. Sometimes, the optimum pre-processing choices are unclear and cannot be validated, given that simulations lag behind the complexity of data. The main idea of this paper is to update the cleaning strategy in CL23, test pre-processing choices, and report the conditions under which  we re-detect the signal. We, therefore, use the CL23 detection as a benchmark and test pipeline features directly on observations. In particular, we want to assess and mitigate the main drawbacks that we have so far experienced with a PCA analysis: signal loss (e.g. in CL23, up to $80\%$ of the signal power was reconstructed a posteriori at the largest scale probed) 
and the ambiguity of the cleaning level choice (which increases the variance linked to the measurement).

We have structured the paper as follows. 
Section\,\ref{sec:data} describes the \hi~intensity mapping observations used in our study. Section\,\ref{sec:theory} presents a mathematical framework for the contaminant subtraction problem and defines our cleaning approach. It also introduces a novel multiscale algorithm, mPCA. 
Section\,\ref{sec:steps} outlines and justifies the data pre-processing steps. Section\,\ref{sec:results} details the results of the various contaminant separations performed. Section\,\ref{sec:model} explains how we compute the power spectra and the theoretical model we assume for their fitting. 
Section\,\ref{sec:crossPk} showcases the cross-correlation of the cleaned intensity maps with the WiggleZ galaxies, constraining the clustering amplitudes, with Table~\ref{table:summary} and Fig.~\ref{fig:results} presenting a summary. 
Finally, Sect.\,\ref{sec:mPCAoutlook}  inspects the cubes cleaned with the new mPCA approach more closely, revealing improved component separation compared to standard PCA.  
We discuss our results and their implications in Sect.\,\ref{sec:discussion} and provide our concluding thoughts in Sect.\,\ref{sec:conclusions}.

\section{The data set}
\label{sec:data}

\begin{figure*}
\centering
    \includegraphics[width=1.8\columnwidth]{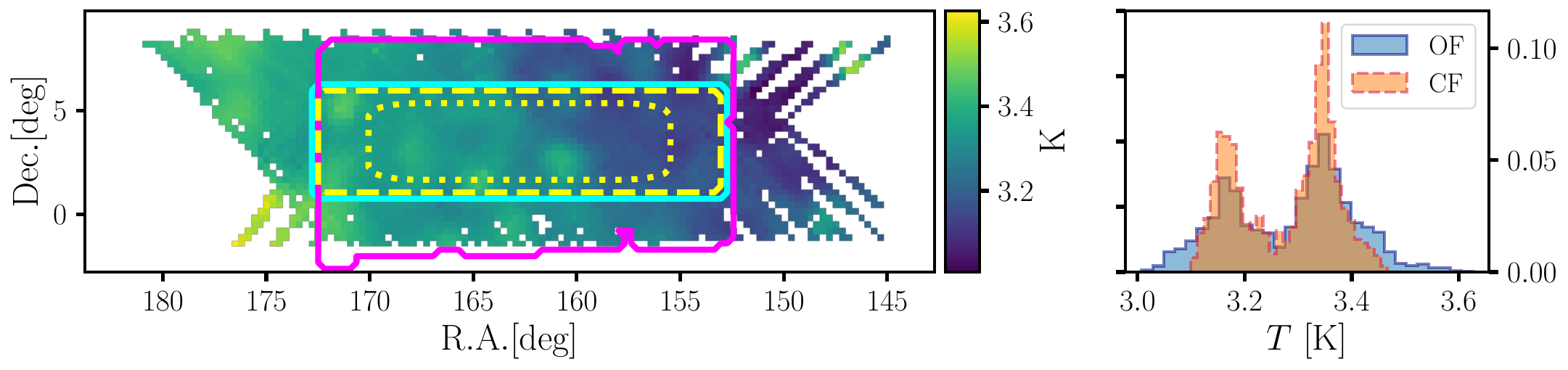}
    \caption{{\it Left panel:} Temperature sky map averaged along the frequency range considered, i.e. $971 < \nu < 1023$ MHz. We highlight the footprint of the WiggleZ galaxies in magenta, the smaller footprint CF where we perform the analysis of this work in cyan, and in yellow the Tukey window function we use for the power spectrum computations (dashed and dotted for the zero and $50\%$ boundaries.).
    {\it Right panel:} Normalised histograms of the sky temperature of the data cubes for the original footprint (OF) in solid blue with respect to the cropped CF in dashed orange. 
    The histograms are computed from the average map for each cube, to marginalise the frequency-dependent evolution of the temperature field. The double-peak structure reflects the galactic synchrotron gradient (low versus high R.A.) present in our sky patch, as shown in the left panel.
    }
    \label{fig:foot}
\end{figure*}

The observations we use belong to the MeerKAT single-dish \hi~intensity mapping pilot survey run in 2019 and have already been described in \citet{wang2021} and used by \citet{Irfan2022,CL23,Irfan2024,Engelbrecht2024}.
Below, we summarise the instrument, observation strategy, and pipeline used to compose the intensity maps.

MeerKAT is a radio telescope in the Upper Karoo desert of South Africa that will eventually become part of the final mid-instrument of the SKAO, SKA-Mid. It comprises 64 dishes of $D_\text{dish} = 13.5$ m in diameter that harbour three receivers (UHF, L, and S bands). Our data were obtained with the L-band receiver ($900-1670$ MHz) using a subset of 250 channels belonging to the range $971 < \nu < 1023$ MHz (corresponding to redshifts $0.39 < z < 0.46$ for the 21-cm emission from \hi),  which is especially free of radio frequency interference (RFI) \citep{wang2021}.

MeerKAT single-dish observations follow a fast scanning strategy at constant elevation to minimise the effect of 1/f noise and retain a continual ground pickup. Dishes move at 5 arcmin s$^{-1}$ along azimuth, scanning about 10 deg in 100 s. The survey targeted a patch of $\sim200$ deg$^2$ in the WiggleZ 11 h field ($153^\circ\,{<}\,\text{R.A.}\,{<}\,172^\circ$ and $-1^\circ\,{<}\,\text{Dec.}\,{<}\,8^\circ$) \citep{Blake:2010xz,Drinkwater:2009sd,WiggleZ:2018def}, away from the galactic plane (where astrophysical foregrounds are brightest) and overlapping with 4031 WiggleZ galaxies, whose positions we cross-correlate with the intensity maps.
Observations occurred during six nights between February and July 2019. After
observing a bright-flux calibrator, each complete scan (i.e. observational block) across the sky patch took about 1.5 hours. We repeated it seven times, resulting in 10.5 hours of combined data. 

\citet{wang2021} detailed the data reduction pipeline for the time-ordered data (TOD) and subsequent map-making. It includes three steps for flagging RFI, bandpass and absolute calibration using known point sources (3C 273, 3C 237, and Pictor A), and the calibration of receiver gain fluctuations based on interleaved signal injection from a noise diode to remove the long-term gain fluctuations due to 1/f noise \citep{Li:2020bcr}. The calibrated total intensity (in Kelvin) TOD are projected into map space (composed by square pixels of $0.3$ deg width) adopting the flat-sky approximation and assuming uncorrelated noise. Finally, the calibrated temperature maps are combined for all scans and dishes at all frequencies by pixel-averaging, yielding the final data cube shown in the left panel of Fig.~\ref{fig:foot}. 

Observing in single-dish mode yields a map with resolution inversely proportional to the dish diameter. 
Specifically, the full-width-half-maximum (FWHM) of the central beam lobe for the channel of frequency, $\nu$, is given by \citep{Matshawule:2020fjz}:
\begin{equation} \label{eq:thetaFWHM}
    \theta_\text{FWHM}(\nu) = \frac{1.16 \, c\nu}{D_\text{dish}}\,,
\end{equation}
with $c$  the speed of light. At our median frequency, it corresponds to $\theta_\text{FWHM} = 1.28$ deg, namely, the size of $3-4$ pixels.

Following CL23, as a result of the map-making and pixel-averaging steps, we estimated the pixel noise variance in the maps, $\hat{n}$, which is inversely proportional to the number of hits (time) stamps. The inverse of the pixel noise
variance is the inverse-variance weight maps, $w_\hi\,{=}\,1/\hat{n}$, that we use later  in our analysis. 

We used the `ABBA' method \citep{wang2021} to remove the sky signal in order to estimate the channel-averaged thermal noise per pixel of the data cube. When considering data within their original footprint, we estimated the noise level to be of about $2.8$ mK; it decreases to $2$ mK when cropping data to a more conservative footprint (Fig.~\ref{fig:foot}, left panel, cyan contour).

\section{Statistical framework}
\label{sec:theory}

We assembled the observed scans so that for each frequency channel, $\nu$, we had a map of the total brightness temperature, $X$. For each given position on the sky (i.e. each pixel, $p$), the observed temperature is the sum of the cosmological 21\,cm signal from \hi~($T_{\hi}$), of the contaminants ($T_{\rm C}$), and of uncorrelated noise ($T_{\rm N}$):
\begin{equation} \label{eq:Tmap}
    X(\nu,p) =  T_{\rm C} (\nu,p) + T_{\hi} (\nu,p) + T_{\rm N}(\nu,p)\,.
\end{equation}
In the separation process, we estimated $T_{\rm C}$ and subtracted it from the original maps to yield the cleaned maps. 
We estimated $T_{\rm C}$ by linearly decomposing it as a sum of $N_{\rm fg}$ foreground components (in the signal-processing literature typically called sources, $S$) modulated by a frequency-dependent amplitude, $A$:
\begin{equation} \label{eq:Tsum}
    T_{\rm C}(\nu,p) = \sum_{i=1}^{N_{\rm fg}} A_i(\nu) S_i(p) \,.
\end{equation}
We compressed all the maps in a data cube, \textbf{X}, with  $N_{\rm pix} \times N_{\rm ch}$ dimensions, with $N_{\rm pix}$ the number of pixels in each map and $N_{\rm ch}$ the number of maps (frequency channels). By merging Eq. (\ref{eq:Tmap}) and (\ref{eq:Tsum}), we can write \textbf{X}  in its matrix form as
\begin{equation} \label{eq:master}
    \textbf{X} = \textbf{A}\textbf{S} + (\textbf{H} + \textbf{N})\,.
\end{equation}
We call \textbf{A} the mixing matrix that regulates the contribution of the $N_{\rm fg}$ components \textbf{S} in the resulting `mixed' signal. Then,  \textbf{H} is the cosmological signal due to the neutral hydrogen 21-cm emission, to be added to a thermal (Gaussian) noise contribution, \textbf{N}. It follows that \textbf{A} has $N_{\rm ch} \times N_{\rm fg}$, while \textbf{S} has $N_{\rm fg} \times N_{\rm pix}$ dimensions. The problem of contaminant subtraction reduces to determining the foreground driven $\hat{\textbf{A}}\hat{\textbf{S}}$ so that \textbf{X}$_{\rm clean}$ = \textbf{X} - $\hat{\textbf{A}}\hat{\textbf{S}}$ can estimate  the cosmological \hi~brightness temperature plus the Gaussian noise contribution as accurately as possible. 
In practice, we only need to determine $\hat{\textbf{A}}$, as getting $\textbf{X}_{\rm clean}$ is equivalent to projecting $\hat{\textbf{A}}$ on the data cube \textbf{X} \citep{carucci2020}:
\begin{equation} \label{eq:Xclean}
   \textbf{X}_{\rm clean} = \textbf{X}-\hat{\textbf{A}}(\hat{\textbf{A}}^\intercal \hat{\textbf{A}})^{-1}\hat{\textbf{A}}^\intercal\textbf{X}   \,;
\end{equation}
in other words, $\hat{\textbf{A}}$ acts as a filter on the data set.

Solving Eq. (\ref{eq:master}) to find the estimate $\hat{\textbf{A}}$ is an ill-posed inverse problem. We need some extra assumptions to move forward. Here, we focus on the Singular Value Decomposition (SVD) and the related principal component analysis (PCA). Here, we  also introduce the multiscale analyses (dubbed mSVD and mPCA), which treat the spatial scales of the maps independently.
 Next, we see how $\hat{\textbf{A}}$ is derived according to the different algorithms used in this work (PCA, SVD, mPCA, and mSVD) and their weighted counterparts (PCAw, SVDw, mPCAw, and mSVDw) when using also the noise information contained in the $w_\hi(\nu,p)$ maps in the separation process.

\subsection{Principal component analysis and singular value decomposition}
\label{sec:SVD}

Overall, PCA is a dimensionality reduction technique that identifies a meaningful basis (in our case, that of the $N_{\rm fg}$  components) to re-express the data set; the principal components should exhibit the fundamental structure of data. Its use is extensive in scope, spanning multitudinous disciplines (see \citet{Jolliffe2016} for a general review). 
PCA assumes that the $\hat{\textbf{A}}\hat{\textbf{S}}$ decomposition maximises decorrelation among the new components, eliminating second-order dependencies. Those new orthogonal components are primarily responsible for the variance in the data, thereby revealing important structural features. It can be demonstrated that the directions corresponding to the largest variances are also associated with higher signal-to-noise ratios \citep{Miranda2008}, which holds true for the contaminants present in our observations.

We can obtain the PCA decomposition as the solution to an eigenvalue-eigenvector problem or equivalent to singular value decomposition (SVD) of the centred data cube since SVD is the generalisation of the eigendecomposition \citep{PCAtutorial}. Indeed, when we are not mean-centring the maps, we say we are performing an SVD data decomposition (in the literature, SVD is also referenced as an `uncentred PCA').

In practice, computing the PCA of the data set $\textbf{X}$ entails (i) subtracting the mean at each frequency to get the centred $\textbf{X}_{\rm ctr}$,
\begin{equation}
    X_{\rm ctr}(\nu_i,p) = X(\nu_i,p)- \frac{1}{N_{\rm pix}} \sum_{j=1}^{N_{\rm pix}} X(\nu_i,p_j)   \,,
\end{equation}
and (ii) computing the eigenvectors of the frequency-frequency covariance matrix. Instead, with SVD, we can go directly to step (ii), without carrying out the mean-centring.
A priori, we do not know whether PCA or SVD is the best approach in our context; both have been used to characterise and remove contaminants from \hi~intensity mapping data sets (e.g. \citet{switzer2013,Anderson2018} used SVD, CL23 used PCA). The statistics literature is not definitive either \citep[e.g. see discussion in][]{Miranda2008}, as the best decomposition depends on the specific problem under consideration; it is common to test different algorithms directly on data and check the solutions \citep[see, e.g. ][]{RemoteSens_example}. 
Here, we also adopted the latter strategy: whether or not to subtract the mean from maps is one of the pre-processing steps we  tested in this work; hence, we refer to the PCA or SVD solutions of the contaminant subtraction procedure throughout the work.

Another common pre-processing choice within the PCA framework in the \hi~intensity mapping context is the weighted PCA (PCAw). This procedure consists of computing the covariance matrix of the weighted data cube, $w_\hi\textbf{X}$, through the inverse noise of the maps 
to downplay the influence of the pixels we are less confident about. As first realised by \citet{switzer2013}, it is convenient not to increase the rank of the data matrix to prevent the frequency structure in the weights from altering the data covariance. Hence, we recast $w_\hi(\nu,p)$ into its 2D projection $w_\hi(p)$, taking the mean along each line of sight. The weights echo the integration time
per pixel; hence, their separability in frequency-pixel is genuine\footnote{Especially considering that, in the narrow frequency band we consider, RFI-flagging has been minimal.}; with the latter projection, we enforced it to be mathematically true.

In the analysis of this work, we aim to test the `weighting' in combination
with the other pre-processing steps and this is why we coupled it to either the PCA
or SVD. 

\paragraph{Summary of PCA-based pipelines.} Schematically, the contaminant subtraction pipeline we followed consists of the following steps.
\begin{enumerate}
    \item We compute the dot product $\textbf{R} = \textbf{D} \textbf{D}^\intercal$, with 
    $$
        \textbf{D} =  \begin{cases}
        \textbf{X}, & \text{for SVD,}\\
        \textbf{X}_{\rm ctr}, & \text{for PCA,}\\
        w_\hi\textbf{X}, & \text{for SVDw,}\\
        w_\hi\textbf{X}_{\rm ctr}, & \text{for PCAw.}
        \end{cases}
    $$

    \item We compute the eigenvectors $\textbf{V}$ of $\textbf{R}$ and order them by their eigenvalues, from largest to smallest.
    \item We use the first $N_{\rm fg}$ eigenvectors (first $N_{\rm fg}$ columns of $\textbf{V}$) to define the mixing matrix $\hat{\textbf{A}}$; namely, $\hat{\textbf{A}}$ is a subset of \textbf{V}: $$\hat{\textbf{A}} = \textbf{V}^{N_{\rm ch} \times N_{\rm fg}}\,.$$
\end{enumerate}
Having defined $\hat{\textbf{A}}$, the solution $\textbf{X}_{\rm clean}$ is obtained through Eq. (\ref{eq:Xclean}).

\subsection{The multiscale approach: Different scales require different treatment}

\begin{figure}
\centering
    \includegraphics[width=0.9\columnwidth]{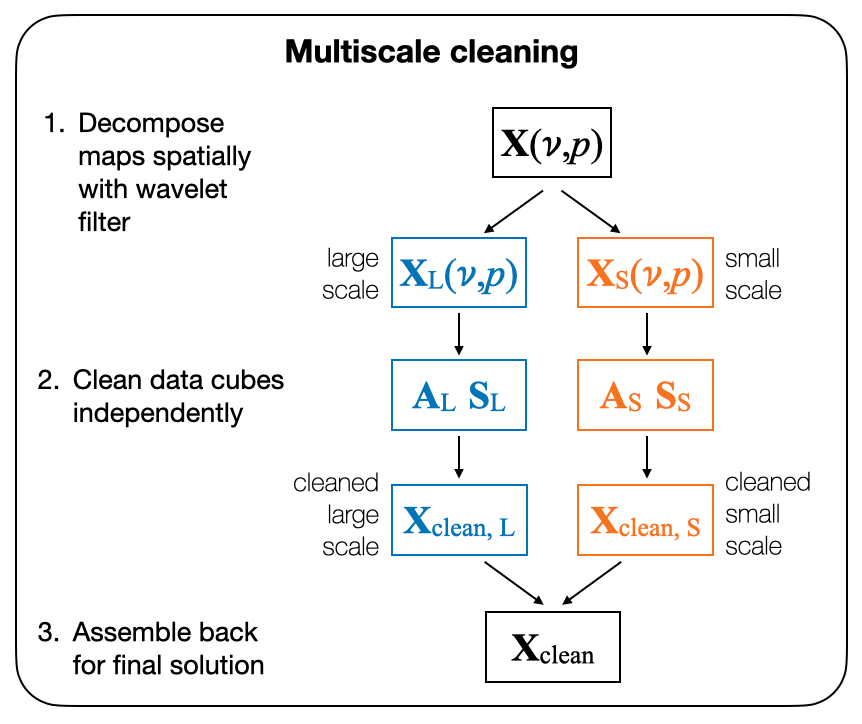}
    \caption{Flowchart describing the multiscale contaminant subtraction.
    }
    \label{fig:flow}
\end{figure}

A typical concern when cleaning \hi~intensity maps with methods based on Eq. (\ref{eq:master}) is that it becomes necessary  to increase the aggressivity of the subtraction (by setting higher $N_{\rm fg}$) to decrease variance in the large spatial scales of the maps. Meanwhile, on small scales, there are  no dramatic changes  during this process, resembling  an `overfitting' condition \citep[e.g. see Fig. 5 in ][]{wolz2017}. Non-observational studies have also begun to spot (and quantify) this trend as more realistic simulations were involved. For instance, from Fig. 15 in \citet{carucci2020}, we can see that a more aggressive cleaning was employed to recover more signal power at large scales at the cost of removing a significant percentage of the signal on the intermediate and small scales. Furthermore, \citet{carucci2020} demonstrated that the wavelet domain is an optimal, multiscale framework for analysing \hi~intensity mapping data, enabling a sparse (compact) representation of the astrophysical foregrounds.

Building upon these previous findings, in this work, we want to test the multiscale cleaning with observations. We treated it as an extra pre-processing choice and, thus, we considered a multiscale PCA (mPCA) and the related mSVD and weighted versions, mPCAw and mSVDw. As we sketch in Fig.~\ref{fig:flow}, once we decomposed the data cubes in different scales, we solved two independent contaminant subtraction problems, entailing different mixing matrices and allowing for different $N_{\rm fg}$.

\begin{figure}
    \includegraphics[width=\columnwidth]{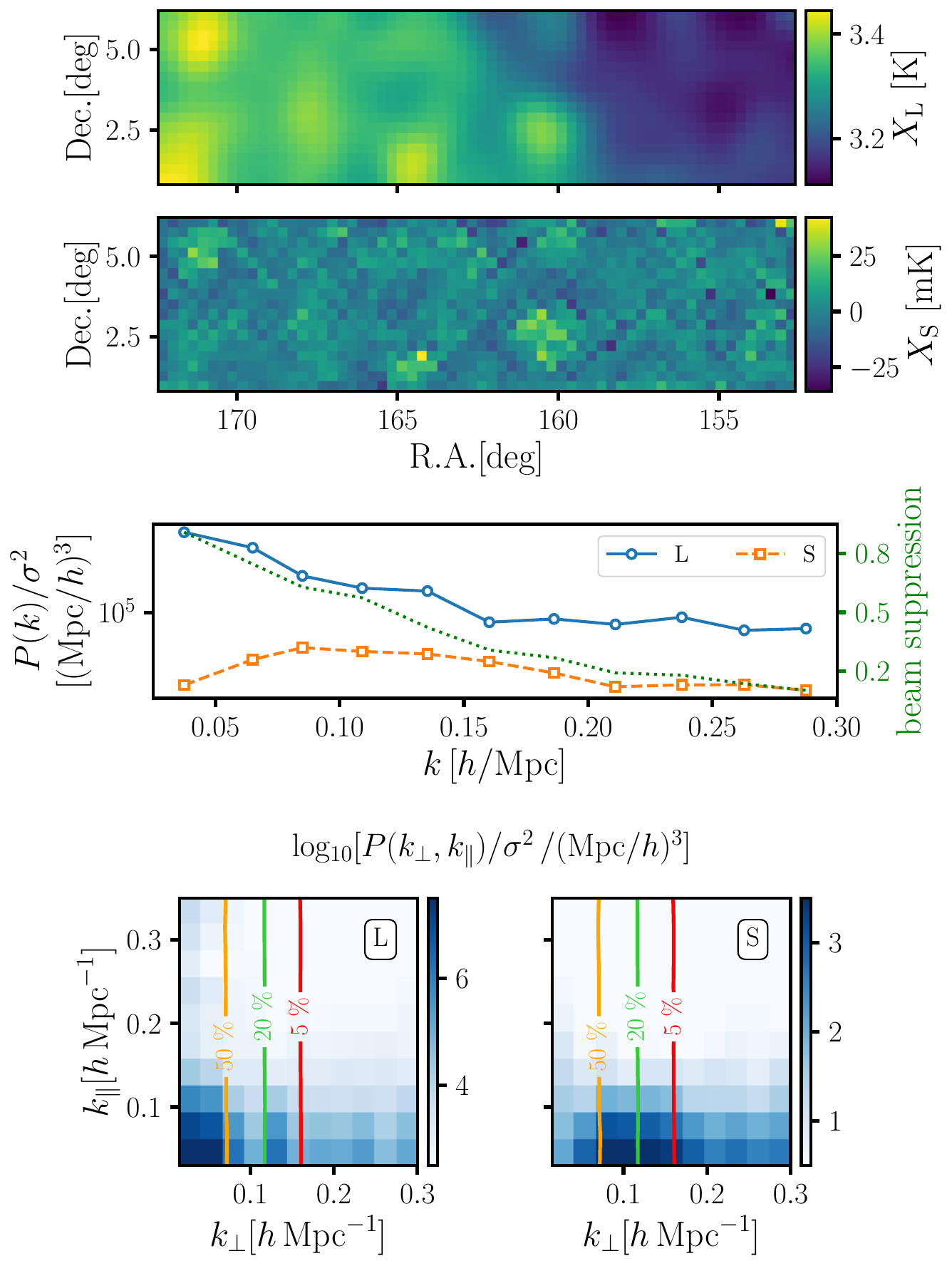}
    \caption{
    {\it First two panels:} 
    Wavelet-filtered large (first panel) and small scale (second) temperature sky map within the conservative footprint (CF) and averaged along all frequencies; i.e. the sum of the two maps above gives precisely the original one (within the cyan contour) in Fig.~\ref{fig:foot}.
    {\it Third panel:} 
     Spherically averaged power spectra of the cubes (large scale in solid blue, small in dashed orange) normalised by the variance of each cube. With a green dotted line, we plot the damping term of the telescope beam (refer to the $y$-axis at the right of the panel). 
    {\it Bottom panels:} 
     Cylindrical power spectra of the large (left panel) and small-scale cubes (right). Since the wavelet filtering is performed in the angular direction, the difference among the resulting cubes is mostly visible along $k_\perp$ ($x$-axis). The beam suppression also acts in this direction: we plot its expected damping term with iso-contour solid lines corresponding to 50, 20, and $5\%$ suppression, from left to right.
    }
    \label{fig:foot_wavelets_pk}
\end{figure}

\paragraph{Starlet filtering.} The wavelet decomposition offers a straightforward framework for analysing multiscale data. Here, we used the isotropic undecimated wavelet, also known as the starlet transform \citep{Starck2007}, which has proven to be well adapted to an efficient description of astrophysical images \citep[e.g.][and, in particular, \citet{carucci2020}  for the single-dish (low-angular-resolution) \hi~intensity maps]{Floer2014,Joseph2016,Offringa2017,Peel2017,Irfan2018,Ajani2020}. 
Starlets have the advantages of (i) representing an exact decomposition; (ii) being non-oscillatory both in real space and in Fourier space, they are efficient at isolating map features represented in either domain; (iii) having compact support in real space, which enables them to prevent systematic effects due to masks and borders.

An intensity map at frequency, $\nu^\prime$, can be decomposed by this transform into a so-called `coarse' version of it, $X_{\rm L}(\nu',p)$, plus several images of the same size at different resolution wavelet scales, $j$,
$$X(p) = X_{\rm L}(p) + \sum_{j=1}^{j_{\rm max}}W_j(p)\,,$$where we dropped the $\nu^\prime$  dependency for brevity. Wavelet-filtered maps, $W_j$, represent the features of the original map at dyadic (powers of two) scales, corresponding to a spatial resolution of the size of $2^j$ pixels, with the largest scale of $X_{\rm L}$ corresponding to the size of $2^{j_{\rm max}+1}$ pixels\footnote{Due to our narrow frequency band, we overlook the frequency dependence of the maps' resolution. To achieve an equal distribution of scales, the wavelet filtering should also be frequency-dependent.}. In our analysis, we set $j_{\rm max} = 1$ after  verifying that using higher wavelet scales does not manifestly improve our results. Hence, applying the starlet decomposition on the observed data cube gives rise to two new cubes, 
\begin{equation} \label{eq:wave}
    X(\nu,p) = X_{\rm L}(\nu,p) + X_{\rm S}(\nu,p)\,,
\end{equation}
which we  refer to as the large-scale and small-scale data sets. 
We display them in the first two panels of Fig.~\ref{fig:foot_wavelets_pk}, averaged along frequency, within the smaller `conservative' footprint (later defined in Sect.\,\ref{sec:steps}): those maps are the wavelet-filtered equivalents of the cyan-contoured map in the left panel of  Fig.~\ref{fig:foot}. 

Since we went on to later analyse data in three-dimensional Fourier space, 
we plotted the power spectra of those cubes in the third panel of Fig.~\ref{fig:foot_wavelets_pk}: the description of `large' and `small' scales holds in $k-$ space, also; although there is no  simple, sharp boundary. 
We compared the behaviour of those three-dimensional spectra with the expected beam suppression with a dotted green line, assuming it is frequency-independent and Gaussian (following Eq. (\ref{eq:thetaFWHM}) and the modelling later described by Eq. (\ref{eq:Pkmodel})). The values of the damping effect of the beam are on the $y$-axis on the right of the panel. The large-scale $X_{\rm L}$ has most of its power for $k\lesssim0.15\,h/$Mpc, where the telescope beam is suppressing up to $70\%$ of the sky signal, then it flattens. The small-scale $X_{\rm S}$ lacks power at small $k$, reaching a peak at $k\approx0.08\,h/$Mpc, corresponding to $\approx30\%$ of beam damping. 
It might be more intuitive to look at the two-dimensional power spectra in the bottom panels of  Fig.~\ref{fig:foot_wavelets_pk} (left for $X_{\rm L}$ and right for $X_{\rm S}$) since the wavelet filtering is applied in angular space. Most of the power of $X_{\rm L}$ is at the largest $k_\perp$ where the beam damping is below $50\%$. The $X_{\rm S}$ power is significant after that $50\%$ bound and slowly decreases.

\paragraph{Summary of the multiscale cleaning.} Once we split the data into large-scale and small-scale cubes, we performed two independent contaminant subtractions, as  described in Sect.\,\ref{sec:SVD}. These subtractions result in two solutions, $\textbf{X}_{\rm L, clean}$ and $\textbf{X}_{\rm S, clean}$, which we summed to get the actual cleaned cube of the multiscale analysis, a unique whole-scale solution, as per Eq. (\ref{eq:wave}). We schematically summarise this process in Fig.~\ref{fig:flow}.

The multiscale framework can be coupled to any cleaning method.
Here, we focus on the PCA-based algorithms already presented. Hence, when we apply PCA (SVD) to both $X_{\rm L, clean}$ with $N_{\rm L}$ components removed and $X_{\rm S, clean}$ with $N_{\rm S}$, we refer to mPCA (mSVD). We can also weight $X_{\rm L}$ and $X_{\rm S}$ with the inverse-variance weights, hence resulting in the weighted counterparts, mPCAw and mSVDw. We notice from Fig.~\ref{fig:foot_wavelets_pk}  that the large scale keeps the dimensionality of the cube, so we can apply an SVD to it (i.e. without subtracting the mean temperature value from the maps). The small-scale maps are inherently zero-centred, so using a SVD on them is equivalent to applying a PCA.

\subsection{General consequences of the component separation}

We expect the cosmological \hi~signal to be far less correlated in frequency and subdominant in amplitude relative to the foregrounds.
These points have two immediate consequences in the separation process. 

First of all, (i) the linear decomposition of \textbf{A}\textbf{S} in Eq. (\ref{eq:master}) is an inconvenient description of the \hi~signal and when forcing that decomposition on data, leakage of some of the \hi~signal into the \textbf{A}\textbf{S} product is unavoidable. Specifically, if, on the one hand, foregrounds can be restricted to a subset of these modes, on the other, the \hi~signal spreads across all modes. It, therefore, becomes crucial to perform the separation task with the smallest possible number $N_{\rm fg}$  of components. For instance, \citet{cunnington2021} report an \hi~signal loss in a PCA analysis of the order $10\%$ and $60\%$ when $N_{\rm fg}$ is set to 5 and 30, respectively, and finds the latter results almost independent of the complexity of the foregrounds considered in their simulation. 
Point (i) holds for any separation method based on Eq. (\ref{eq:master}), although different algorithms may lead to different amounts of signal loss.

Furthermore, (ii) the cosmological signal is inherently coupled to the uncorrelated noise component---such as the thermal part of the instrumental noise---hence, we do not attempt to separate $\textbf{H}+\textbf{N}$ at the cleaning stage\footnote{One possibility is to remove noise after the cleaning by cross-correlating data subsets, since data splits (e.g. in time) contain the same cosmological signal but uncorrelated noise contributions  \citep[][]{switzer2013,wolz2017,Anderson2018,wolz2022}.
}. In what follows, with `recovered signal', `cleaned data cube', and similar expressions, we mean the estimate of the cosmological \hi~signal plus the uncorrelated component of the noise, $\textbf{X}_{\rm clean}$.

\paragraph{Considering the number of components $N_{\rm fg}$.}

\begin{figure}
\centering
    \includegraphics[width=0.8\columnwidth]{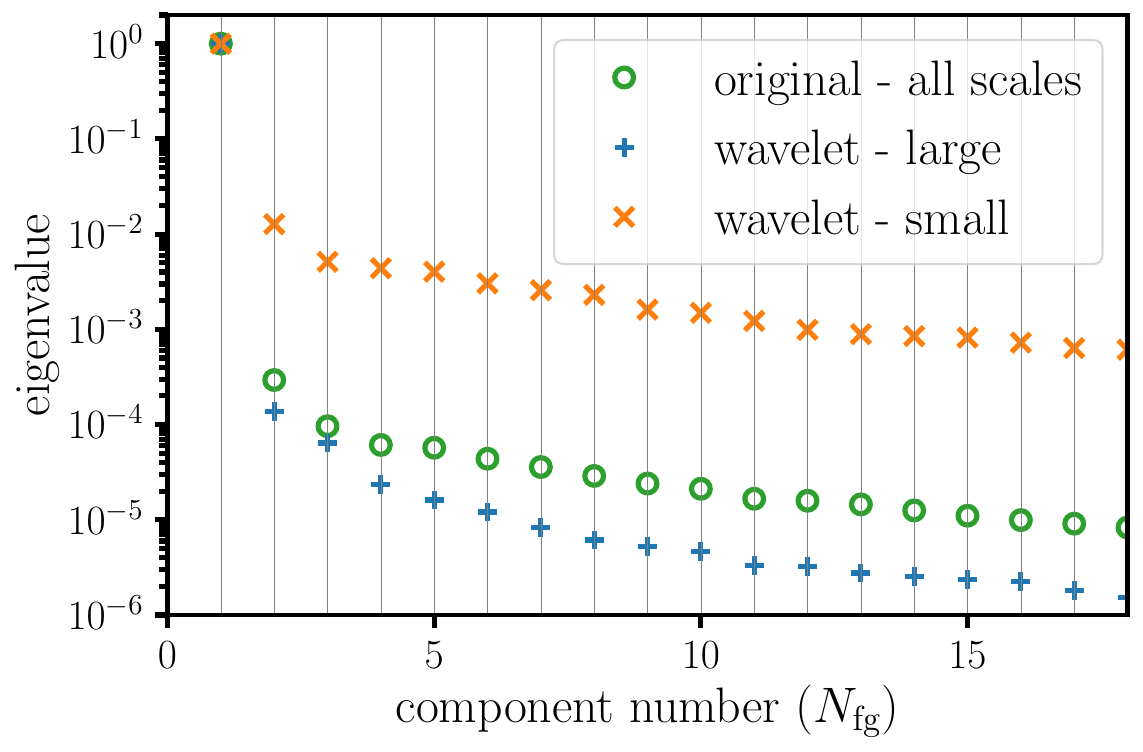}
    \caption{Normalised eigenvalues of the frequency-frequency covariance matrix of the data cube. 
    Circles refer to the original cube, plus signs to the wavelet-filtered large-scale cube, and `x' crosses for the small-scale cube. The large-scale eigenvalues drop down faster with $N_{\rm fg}$: the PCA assumption holds better in this case, and a few modes are enough to describe the large-scale data set. These values correspond to mean-centred maps cropped on the conservative footprint, although the eigenvalues behaviour does not change significantly for the other cases.
    } 
    \label{fig:eigen}
\end{figure}

PCA-like methods are dubbed `blind' or `unsupervised': no prior information on the cosmological signal is required. However, we do not know a priori the optimal number of components $N_{\rm fg}$ needed to describe the contaminants and preserve as much cosmological signal as possible.

Typically, as an exploratory analysis, we can plot the eigenvalues of the covariance eigendecomposition in decreasing order, as we do in Fig.~\ref{fig:eigen}. The number of plotted values before the last substantial drop in the magnitude of eigenvalues suggests how many components are needed to `explain' the data, that is to say, to account for the largest part of the data variance. Indeed, in the contexts where PCA is successful, the change in magnitude among the eigenvalues is unquestionably visible (in the literature, these plots are called `scree plots'). Works based on simulations show evident scree plots; the scree becomes less and less pronounced when additional systematics are modelled on top of astrophysical foregrounds \citep[e.g.][]{carucci2020}, highlighting the mode-mixing at play. Setting $N_{\rm fg}$ unambiguously in observations is challenging. Methods that automatically find the appropriate sub-space where contaminants live typically need prior information, for instance, on the signal covariance \citep{Shaw2014,olivari2016}.

Here, we reverse the argument and claim that if different pre-processing or filtering choices lead to various `scree' eigenvalues behaviour, those that emphasise the magnitude drop are scenarios where the PCA assumptions hold better, resulting into an efficient decomposition. For instance, looking at Fig.~\ref{fig:eigen}, 
the large-scale cube is the data set for which the PCA assumptions hold better (blue plus signs); hence, we can anticipate that on the large scale, PCA will be more efficient---a small $N_{\rm L}$ will be enough to describe contaminants---compared to the small-scale counterpart and the PCA without any wavelet filtering.
For each cleaning method, we  determined the values of $N_{\rm fg}$ (or $N_{\rm L}$, $N_{\rm S}$) by identifying those that yield the best cross-correlation measurements with the external galactic data set.

\section{Pre-processing steps}
\label{sec:steps}

Here, we discuss the pre-cleaning choices we made to optimise the contaminants-signal separation, highlighting the differences in comparison  to the CL23 analysis we have built upon.

\begin{figure*}
\centering
    \includegraphics[width=2\columnwidth]{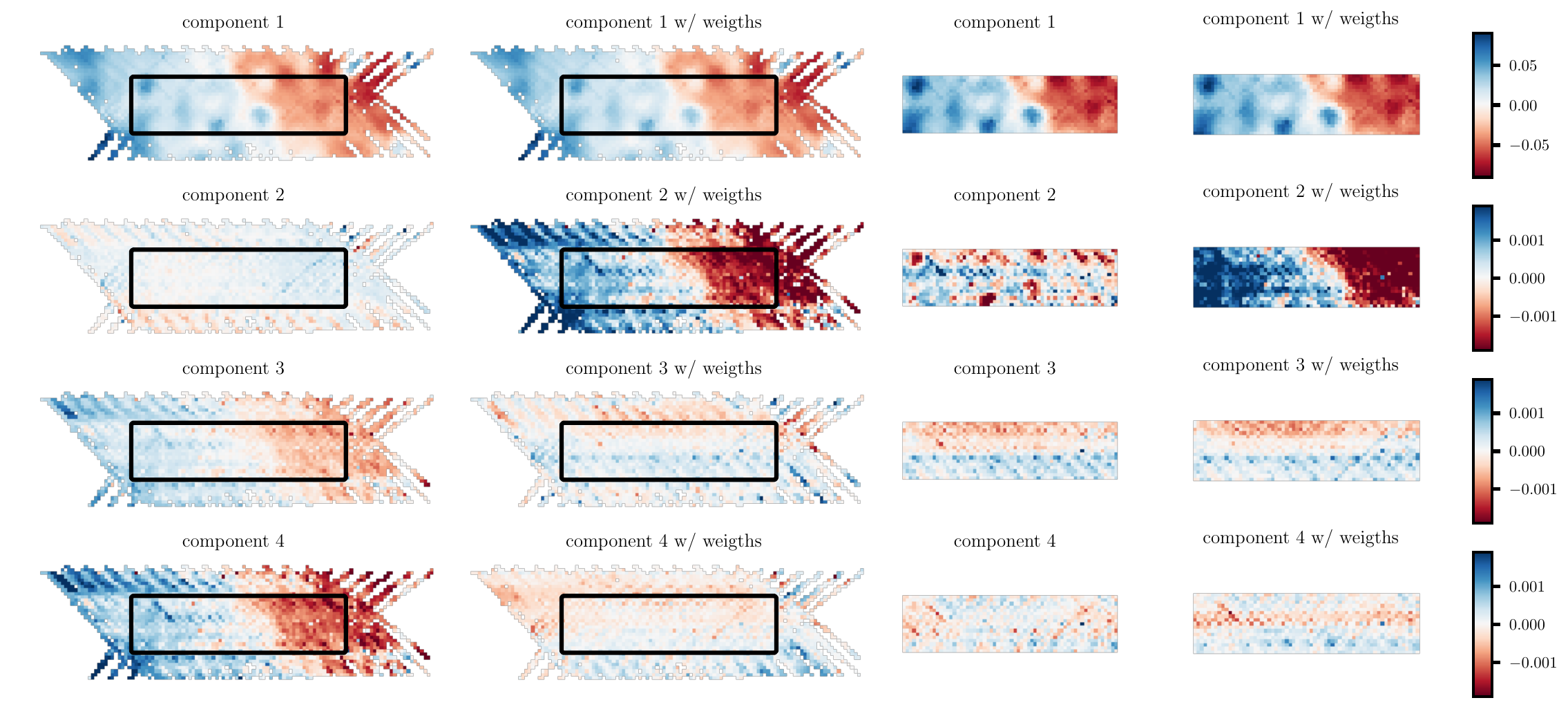}
    \caption{First four components $\hat{\textbf{S}}_{1i}$ to $\hat{\textbf{S}}_{4i}$ (from top to bottom) in the case where we apply PCA on the original footprint OF (first column) or on the conservative footprint CF (third column) and in the cases where we apply a weighted PCAw (second and fourth columns). All plotted maps are normalised to highlight each component's spatial features and pixel variance. Component-number (row) wise, we impose the same range of values as per the colour bars on the right. Some sources get saturated, highlighting the higher pixel variance they capture than the unsaturated counterparts (especially for the second component). The black contours in the OF maps in the first two columns mark the CF boundaries. 
    }
    \label{fig:sources}
\end{figure*}
\paragraph{PCA-informed footprint.} In the left panel of Fig.~\ref{fig:foot}, we show the original footprint, OF, of the data set. With a solid cyan line, we highlight the smaller `conservative' footprint, CF. We have defined the CF by analysing the first four components $\hat{\textbf{S}}$ found by PCA, projecting $\hat{\textbf{A}}$ on $\textbf{X}$, for the OF analysis; we show these first four components in the first column of Fig.~\ref{fig:sources}. We have drawn iso-contours in normalised intensity in the component maps and found that most of the pixel variance captured by the PCA decomposition was contained in the pixels outside  the CF, marked by a solid black boundary. The PCA framework is known to be prone to outlier mishandling\footnote{PCA decomposes data minimising 
the sum of the squares of deviations from the mean value.  
The squaring from the outliers dominates the total norm and, therefore, drives the `principal' components of PCA.}, hence the need to suppress their influence either by removing them from the data set, adopting the CF, or by performing a weighted PCAw since the outlier pixels correspond to those with higher noise\footnote{See bottom map of Figure 15 in \citet{wang2021}, that shows the total number of time samples that were combined in each pixel, i.e. inversely proportional to the weights of PCAw and SVDw.}. A close look at the second column of Fig.~\ref{fig:sources} suggests that the weighting scheme might not be sufficient and that the complete pixel-flagging choices (last two columns) deliver more reasonable principal components (see later in Sect.\,\ref{sec:results} a more in-depth discussion on what we mean by a `reasonable' morphology of the sources).
We later compare the two options more quantitatively, weighting versus cropping.

We show the temperature histograms of the data cubes in the right panel of Fig.~\ref{fig:foot}. The pixels that have been cut out with the new footprint, after looking at the PCA-derived components, are indeed outliers of the original data temperature distribution.
\paragraph{No extra channel flagging.} CL23 performed an extra flagging round on the channels displaying high variance peaks before the cleaning. After channels have been discarded, either we inpaint the missing maps, or we ignore their absence and perform the cleaning on the chopped data cube. CL23 did the latter. However, the PCA decomposition boils down to recovering templates that, modulated in frequency, sum up to the full data cube. When the modulation in frequency is spoiled by the missing channels, the decomposition becomes trickier; in other words, it becomes harder for component analyses of this kind to determine mixing matrices with discontinuities, as shown in \citet{carucci2020,soares2022}. Here, we chose to keep all the 250 channels in the analysis. In Appendix\,\ref{app:test}, we checked whether extra flagging could have been beneficial in the set-up of this analysis, ultimately finding it to be detrimental, in agreement with  simulation-based works.

Additionally, retaining the full frequency range in the analysis allows us to compensate for the information lost with the CF cropping: we worked with a comparable number of voxels as in CL23, however, with a lower estimated noise (Sect.\,\ref{sec:data}).
\paragraph{No extra smoothing of the maps.} Cleaning algorithms based on Eq. (\ref{eq:master}) perform better when maps share a common resolution. Indeed, in the decomposition, the templates $\hat{\textbf{S}}$ get modulated in frequency through $\hat{\textbf{A}}$; however, the mixing matrix $\hat{\textbf{A}}$ cannot accommodate for resolution differences at the map level, as it gives only an overall amplitude adjustment to the $\hat{\textbf{S}}$ components in the frequency direction. Given this intrinsic limitation of the $\hat{\textbf{A}} \hat{\textbf{S}}$ decomposition, accurate beam deconvolution should be ideally performed before the cleaning. CL23 opted for smoothing all maps with a frequency-dependent Gaussian kernel to achieve a common resolution $1.2$ times worse than the largest resolution (i.e. with a FWHM that is $1.2$ times greater than that of the lowest frequency map, assuming it is Gaussian). 
CL23 were not the first to opt for the `reconvolution' choice \citep[e.g.][]{masui2013,wolz2017,Anderson2018,wolz2022}, and, besides downgrading the maps to a common resolution, the extra smoothing is essentially a low-pass filter that removes the small-scale information in the maps, which one can relate to spurious artifacts. However, the effects of the channel-dependent Gaussian smoothing on the properties of the real-beam-convolved \hi~field have not been fully understood yet. \citet{spinelli2022} performed different cleaning methods on simulated data with and without the frequency-dependent smoothing and found the latter practice to be detrimental in all the cases analysed 
(see also \citet{Matshawule:2020fjz} for the effects of a realistic beam in the cleaning process).  
Here, we opted for the simplest option: leave the maps' resolutions as they are. Moreover, the beam resolution does not change dramatically within the frequency range we use ($1.24 - 1.31$ deg), and the level of uncertainties in our measurement allows us to nevertheless model the theoretical expectation for the three-dimensional power spectrum with a Gaussian beam relative to the median frequency of the data cube (Sect.\,\ref{sec:model}).

We  check in Appendix\,\ref{app:test} what reconvolution could do in terms of the final measurements. We find no improvements and even a detrimental effect in the case of the multiscale cleaning.

\section{Results of the component separation}
\label{sec:results}

\subsection{Performing the cleaning within the conservative footprint}

\subsubsection{Mixing matrix}

\begin{figure}
    \includegraphics[width=0.9\columnwidth]{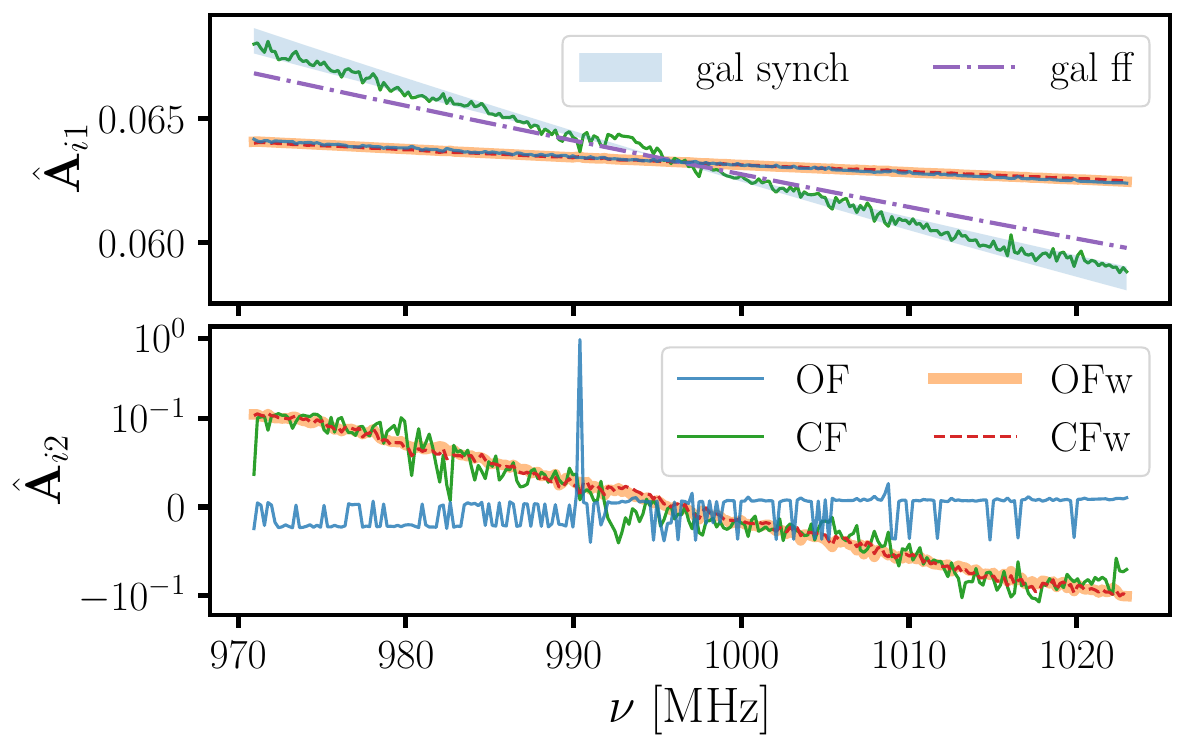}
    \caption{
    Normalised first two modes derived by the PCA analysis performed on the original footprint OF (solid blue line) and the conservative footprint CF (solid orange), and their weighted version `OFw' (solid green) and `CFw' (red dashed). For comparison, in the first mode plot (top panel), we show a proxy for the spectral index of the galactic free-free emission (dash-dotted, spectral index of $-2.13$, \citet{Dickinson2003}) and a range of possible galactic synchrotron spectral indexes (blue shaded area, between $-3.2$ and $-2.6$, \citet{Irfan2022}). 
    } \label{fig:mixmat_foot}
\end{figure}

We started by applying the PCA cleaning (as per the algorithm described in Sect.\,\ref{sec:SVD}) in the OF and CF.
The mixing matrix has all the information for getting to the PCA solution; hence, we started by looking at the derived mixing matrices. 
In Fig.~\ref{fig:mixmat_foot}, we show the first two modes (components at the top for the first and bottom panel for the second) of the PCA-like decompositions. 
Especially for the PCA analysis within the OF (blue solid lines), the second principal mode $\hat{\textbf{A}}_{i2}$ (bottom panel) is completely dominated by sharp peaks at specific channels: the PCA decomposition uses this mode to describe the large variance fluctuations in a subset of the maps. In contrast, the behaviour of $\hat{\textbf{A}}_{i2}$ gets closer to the expected power-law-like when we take weights into account with PCAw (solid orange line). When analysing the CF (green and red lines), modes 1 and 2 are also spectrally smooth, as the galactic diffuse emissions that should dominate the maps. 

While there is no one-to-one relationship between PCA-derived components and actual astrophysical objects, it is noteworthy that the first mode of PCA, when run on the CF, exhibits a frequency behaviour that closely resembles the diffuse synchrotron emission of the Milky Way \citep{Irfan2022}. However, a principal component doing better at isolating the galactic synchrotron does not necessarily indicate that the residuals are cleaner; a first principal component that combines, e.g. free-free and synchrotron emissions, would also reflect the expected dominant contributions in the data. Indeed, all first modes displayed in the top panel of  Fig.~\ref{fig:mixmat_foot}  `have something to do' with the galaxy, which is expected and reassuring.

The SVD  mixing matrices (with or without weighting and for the different footprints) are extremely similar to the PCA counterparts; hence, we do not show them here.

\subsubsection{Sources}

The modes (i.e. the columns of $\hat{\textbf{A}}$) described above are associated with the corresponding components: $$\hat{\textbf{S}}= (\hat{\textbf{A}}^\intercal \hat{\textbf{A}})^{-1}\hat{\textbf{A}}^\intercal\textbf{X} \,.$$We show the first four in Fig.~\ref{fig:sources}, from top to bottom; the first two columns refer to the analysis with the OF, last two columns to the CF, without and with weighting. Components associated with the OF analysis are different than those associated with the CF. Even though the PCA components are not straightforward to interpret as they cannot be exactly paired to physical emissions, they are nevertheless informative---as for their associated frequency behaviour described above. For example, can we see the galactic morphology influence in them? 
For the OF case, components 1, 3 and 4 display the left-to-right gradient that is linked to the synchrotron at this sky position \citep{Irfan2022}, whereas, for the PCA solution for the CF, the synchrotron gradient is visually present for the first component only. The other three likely mainly capture the morphology of some non-astrophysical systematics, given e.g. the sharp horizontal boundaries\footnote{In \citet{MK_GAMA}, with further L-Band observations, we have noticed that these horizontal stripes are linked to RFI contamination during the downlink of a 50 km away mobile communication tower detected by the beam side-lobes. The development of a method to deal with this effect before the cleaning stage is ongoing.} of component 3 and the zigzag structure 
of component 4. In the CF case, components that look visually galactic are linked to smooth modes (see above and in Fig.~\ref{fig:mixmat_foot}). In contrast, the situation is more confusing for the OF case: galactic-like components are associated with non-smooth modes, too. Indeed, when working with the OF, we are witnessing more `mode-mixing'.

Again, we do not show SVD results for brevity.  
The SVD solutions appear somewhat different regarding the $\hat{\textbf{A}}\hat{\textbf{S}}$ decomposition; however, these differences are not appreciable enough to make claims at this stage of whether the SVD analysis is more or less optimal than PCA (e.g. the pre-processing choice of using the CF rather than the OF has a much more significant impact). We let the cross-correlation measurements quantify the PCA-SVD distinctness.

\subsubsection{Variance}

\begin{figure}
\centering
    \includegraphics[width=0.9\columnwidth]{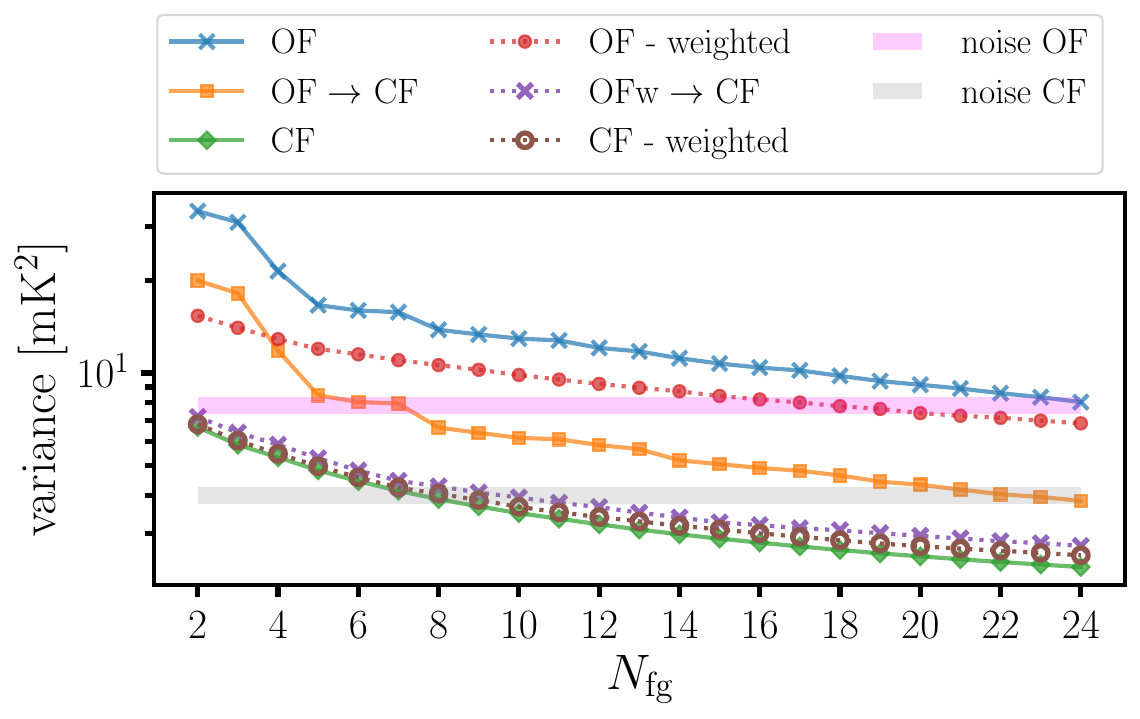}
    \caption{Variance in the PCA-cleaned cubes as a function of 
    $N_{\rm fg}$. The solid blue line corresponds to a PCA cleaning applied to the original footprint OF that, if cropped a-posteriori on the smaller CF, corresponds to the variance displayed by the solid orange line. The solid green line corresponds to PCA performed directly on the CF. We compare these results with a weighted PCAw with dotted lines, in the OF case (red), OF-later-cropped (violet) and CF (brown).  
    The horizontal stripes mark the estimated noise level of the OF (pink) and CF (grey) cubes. 
    These results do not change for an SVD analysis.
    } 
    \label{fig:variance_foot}
\end{figure}

We expect the variance to decrease as a function of $N_{\rm fg}$. Furthermore, to minimise cosmological signal loss that increases with $N_{\rm fg}$, we would aim to reach the expected noise level of the data with the smallest $N_{\rm fg}$ possible. We recall here that noise dominates the cosmological signal in this data set. Hence, even though this exercise cannot tell us how much residual contamination is still present in the cubes, we can compare the PCA-based pipelines as the faster (i.e. lower $N_{\rm fg}$), they reach the noise level, the more optimal. We do so with the solid lines in Fig.~\ref{fig:variance_foot}.  
The blue line, corresponding to a PCA analysis on the original footprint OF, reaches the OF noise level (pink stripe) at $N_{\rm fg} = 23-24$; whereas the green line (PCA on the CF) reaches the CF noise level (grey) at $N_{\rm fg} = 7-8$. Performing the PCA decomposition on the CF leads to a more efficient signal component separation. 

By working on the CF, we are suppressing the high variance pixels at the beginning of our analysis, namely, before the cleaning; thus, we might want to consider whether: (i) the previous result is due to the missing outlier pixels rather than a more efficient cleaning and (ii)  performing a weighted PCAw on the full OF  could allow us to suppress the influence of the high-variance pixels in the signal separation, without cropping the maps. We tested both hypotheses and checked whether they lead to a cleaning efficiency comparable to (or even better than) the PCA applied on the CF.

Regarding hypothesis (i), we cropped the cleaned cubes obtained with the PCA on the OF to the CF, by trimming the maps after the cleaning. The results are shown in orange in Fig.~\ref{fig:variance_foot}. We note that we are not improving the OF results, as we keep reaching the noise (in grey for the CF) around $N_{\rm fg} = 21 - 22$. 
The extra channel flagging is a condition for PCA to deliver a better decomposition. The filtering solution differs from the OF case for the retained pixels, as previously checked by looking at the modes and components of the decomposition.

To check hypothesis (ii), we employed PCAw. The results are shown by the red dotted line in Fig.~\ref{fig:variance_foot}: we improve upon the plain PCA solution (blue line) by being more effective and reaching the noise floor at $N_{\rm fg} = 15-16$, but not as efficiently as in the CF case (green line). Instead, performing a weighted PCAw for the CF case (dotted violet line) does not improve upon the plain PCA. The weight map inside the CF is indeed more uniform, making it less worthwhile to apply the weighting.

We have demonstrated that if data do not have an approximately uniform weight across the cube, then a weighted decomposition is worthwhile in the PCA framework, although it is not as crucial as removing the outlier pixels in the temperature distribution altogether.

Summarising this section, we have found that a PCA analysis on the small CF is more effective than the OF and than its weighted version. Similar conclusions apply to SVD.
We continued working solely with the CF in the remainder of this work.

\subsection{Cleaning spatial scales independently}

In this section, we take a closer look at mixing matrices, sources, and variance of the multiscale cleaning solutions. We focus mostly on mPCA, as we find no appreciable differences with respect to mSVD.

\subsubsection{Mixing matrix}

\begin{figure}
\centering
    \includegraphics[width=0.9\columnwidth]{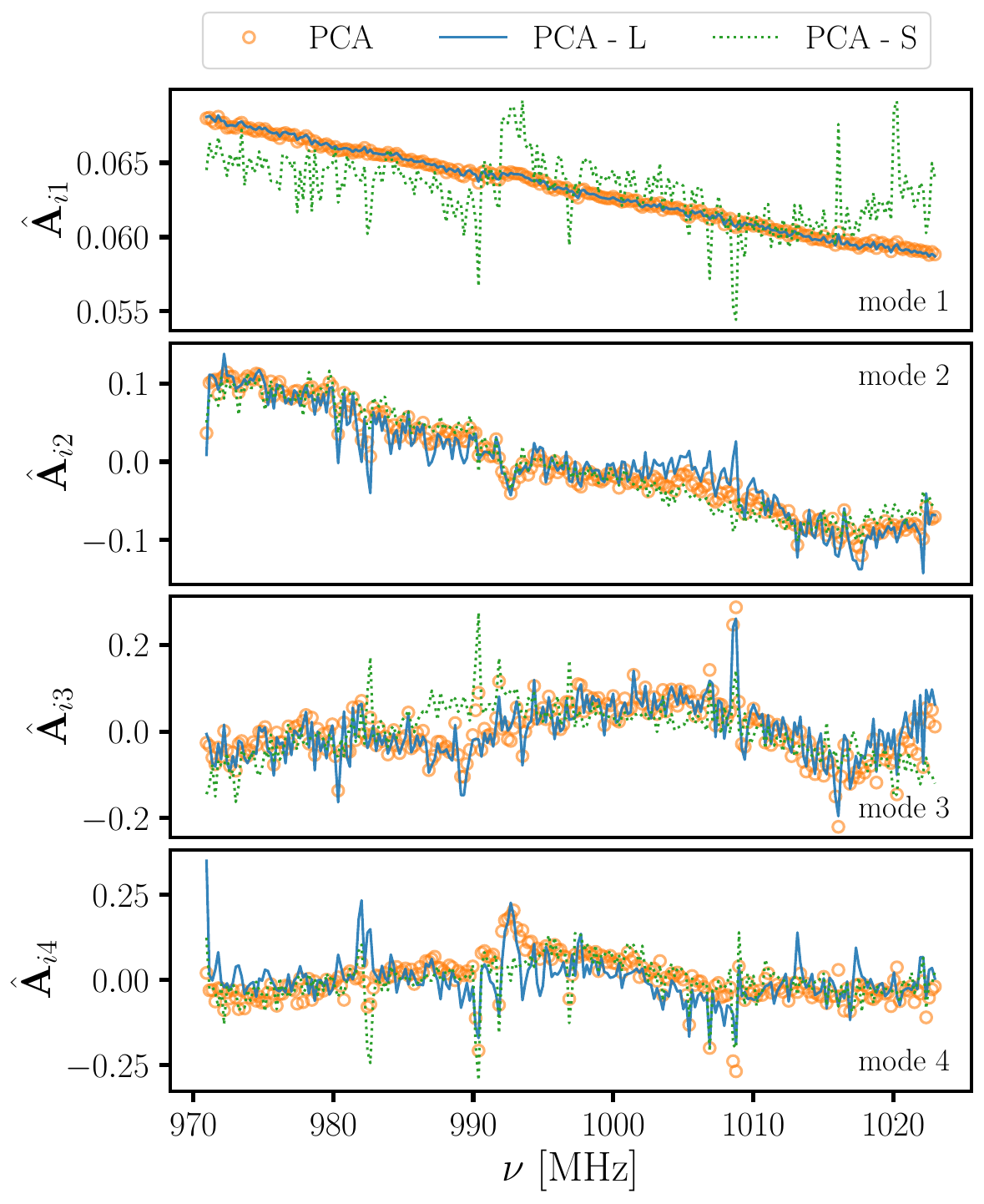}
    \caption{Normalised first four modes (from the top to the bottom panel) derived by the PCA analysis (empty orange circles) and the multiscale mPCA (lines). With mPCA, we derive two mixing matrices, $\hat{\textbf{A}}_{\rm L}$ for the large and $\hat{\textbf{A}}_{\rm S}$ for the small scales, and we show their first modes with solid blue and dotted green lines respectively. The first two modes of all three mixing matrices look  correlated, as the main astrophysical foregrounds are driving the first modes at large and small scales alike. Nevertheless, the extra fluctuations of the $\hat{\textbf{A}}_{\rm S}$ first mode is a hint that the decomposition at the small-scale is characterising extra contributions more relevant at this spatial scale than at the large one.
    } 
    \label{fig:mixmat_mPCA}
\end{figure}

Overall, mPCA corresponds effectively to two PCA analyses, PCA-L and PCA-S, performed independently on the maps' large and small spatial scales and acting on them separately. We started by looking at the two mixing matrices, $\hat{\textbf{A}}_{\rm L}$ and $\hat{\textbf{A}}_{\rm S}$, and see which modes they will project out from the data. We plot the first four modes of both matrices in Fig.~\ref{fig:mixmat_mPCA}, solid blue and dotted green for $\hat{\textbf{A}}_{\rm L}$ and $\hat{\textbf{A}}_{\rm S}$ respectively, and compare them with the plain PCA solution too (empty orange circles). Each panel from top to bottom refers to a mode. The first two modes are almost identical for PCA and PCA-L; instead, the last two differ mainly in the position of the peaks. Since these modes are projected on the entire data cube for PCA and only on the coarse scale of the maps for PCA-L, this finding implies that, in the case of the plain PCA, the large-scale information drives the decomposition, imposing its solution (filtering) also on the small scale of the maps. Indeed, the first two modes of the small-scale driven PCA-S follow the same trend but are much more fluctuating: they are accommodating for channel amplitude fluctuations of these first modes occurring just on the small scale.

\subsubsection{Sources}

\begin{figure*}
\centering
    \includegraphics[width=2\columnwidth]{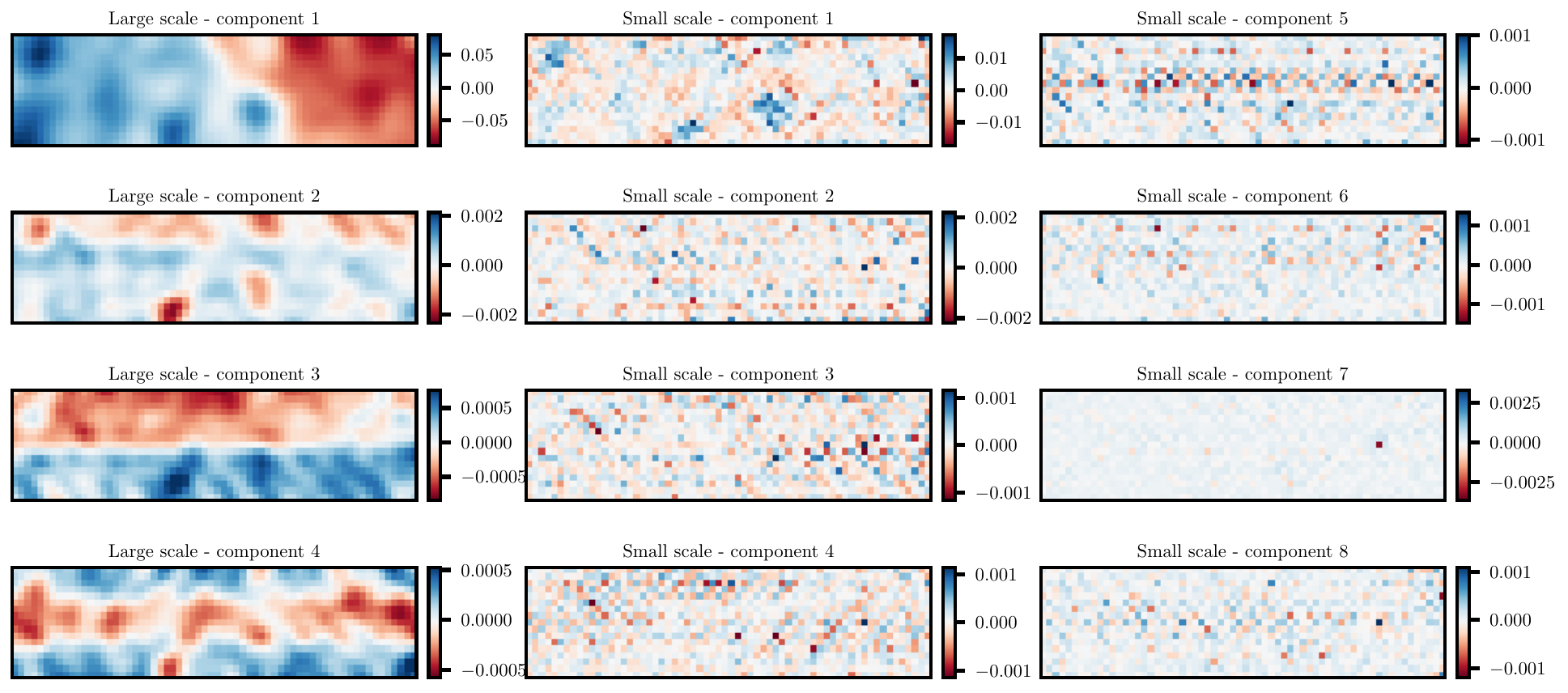}
    \caption{
    First four normalised components $\hat{\textbf{S}}_{\rm L}$ in the case we apply PCA on the large-scale only (left column) and the first eight $\hat{\textbf{S}}_{\rm S}$ on  a small scale. Large-scale components (left column) reflect the coarse structure of the whole-scale PCA decomposition (third column of Fig.~\ref{fig:sources}), whereas, the small-scale components, except the first one to a little extent, do not show any recognisable astrophysical `morphology' in them: they characterise other, small-scale, features, equally crucial in the PCA decomposition (those are the first eight principal components).
    }
    \label{fig:sources_mPCA}
\end{figure*}

The modes discussed above correspond to the components $\hat{\textbf{S}}_{\rm L}$ and $\hat{\textbf{S}}_{\rm S}$ that we show in Fig.~\ref{fig:sources_mPCA}, where the left column is for the large-scale analysis and the rest for the small-scale. 
We can compare these maps with the whole-scale $\hat{\textbf{S}}$ components shown in the third column of Fig.~\ref{fig:sources}. The $\hat{\textbf{S}}_{\rm L}$ maps indeed reflect the large scale of $\textbf{S}$, namely, they are coarse-grained versions of the $\textbf{S}$ maps. Again, this confirms the hypotheses that (i) the large-scale information drives the plain PCA analysis when working with the whole-scale original cube; (ii) the smooth astrophysical emissions acting as foregrounds correspond to morphologically wide, large-scale regions, compared to the sky area and the intrinsic spatial resolution of the observed maps.

It is harder to interpret the $\hat{\textbf{S}}_{\rm S}$. They are indeed descriptions of the small scale decomposition, breaking down the `stripes' and `zigzag' seen in $\hat{\textbf{S}}$; except maybe the first $\hat{\textbf{S}}_{\rm S}$ component, where we do see hints of the point sources present in our footprint \citep{wang2021}. It is indeed this first $\textbf{S}_{\rm S}$ that corresponds to a mode with a similar trend with what is seen in the mixing matrices $\hat{\textbf{A}}$ and $\hat{\textbf{A}}_{\rm L}$: we expect it to carry some information of the astrophysical foregrounds. To a lesser extent, this is true also for the second $\hat{\textbf{S}}_{\rm S}$ component: it is associated with a smooth mode, close to the second mode of $\hat{\textbf{A}}$ and $\hat{\textbf{A}}_{\rm L}$; yet in this case we cannot recognise an astrophysically motivated morphology. Probably, the latter is because these astrophysical emissions have leaked in a systematics-driven component. 
For concision, we do not show the modes 5 to 8 associated with those last four $\hat{\textbf{S}}_{\rm S}$ components: they are spiky and oscillating around zero (modes with short correlation in frequency), so it is hard to relate them to any motivated signal component or systematic contaminant. We might ask whether these small-scale fluctuating components are actual cosmological signal. Even though we always incur some signal loss, these are still the first eight principal components, those of highest variance, and typically smoother than components that would be next in the decomposition. The PCA-S is no more and no less harmful in terms of signal loss than PCA-L. It is just acting on smaller scales, as we explicitly check in Sect.\,\ref{sec:sims}, where we test on simulations the filtering effect of mPCA.

\subsubsection{Variance}

\begin{figure}
\centering
    \includegraphics[width=0.9\columnwidth]{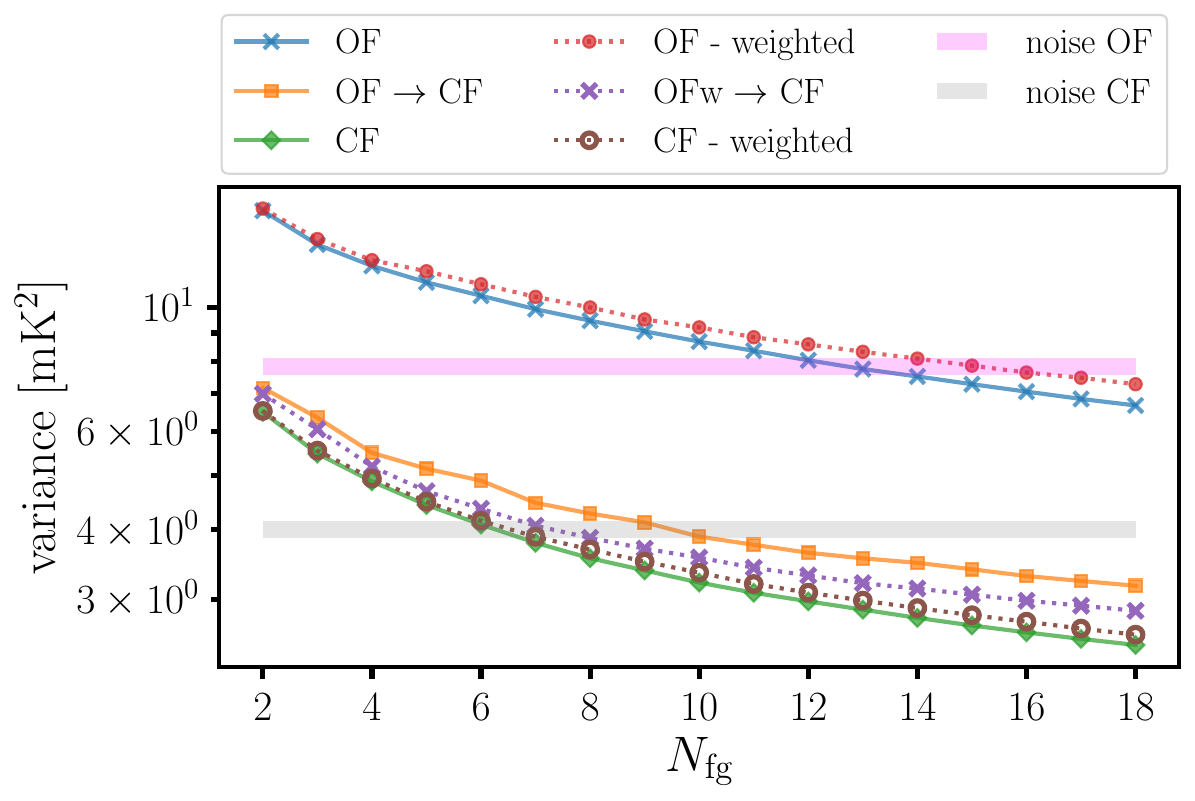}
    \caption{Variance in the mPCA cleaned cubes as a function of the number of components removed when setting $N_{\rm L}=N_{\rm S}$ ($= N_{\rm fg}$, as in the $x$-axis). The legend is the same as in Fig.~\ref{fig:variance_foot}. These results do not change for an mSVD analysis.
    } 
    \label{fig:variance_mSVD}
\end{figure}

\begin{figure}
    \includegraphics[width=\columnwidth]{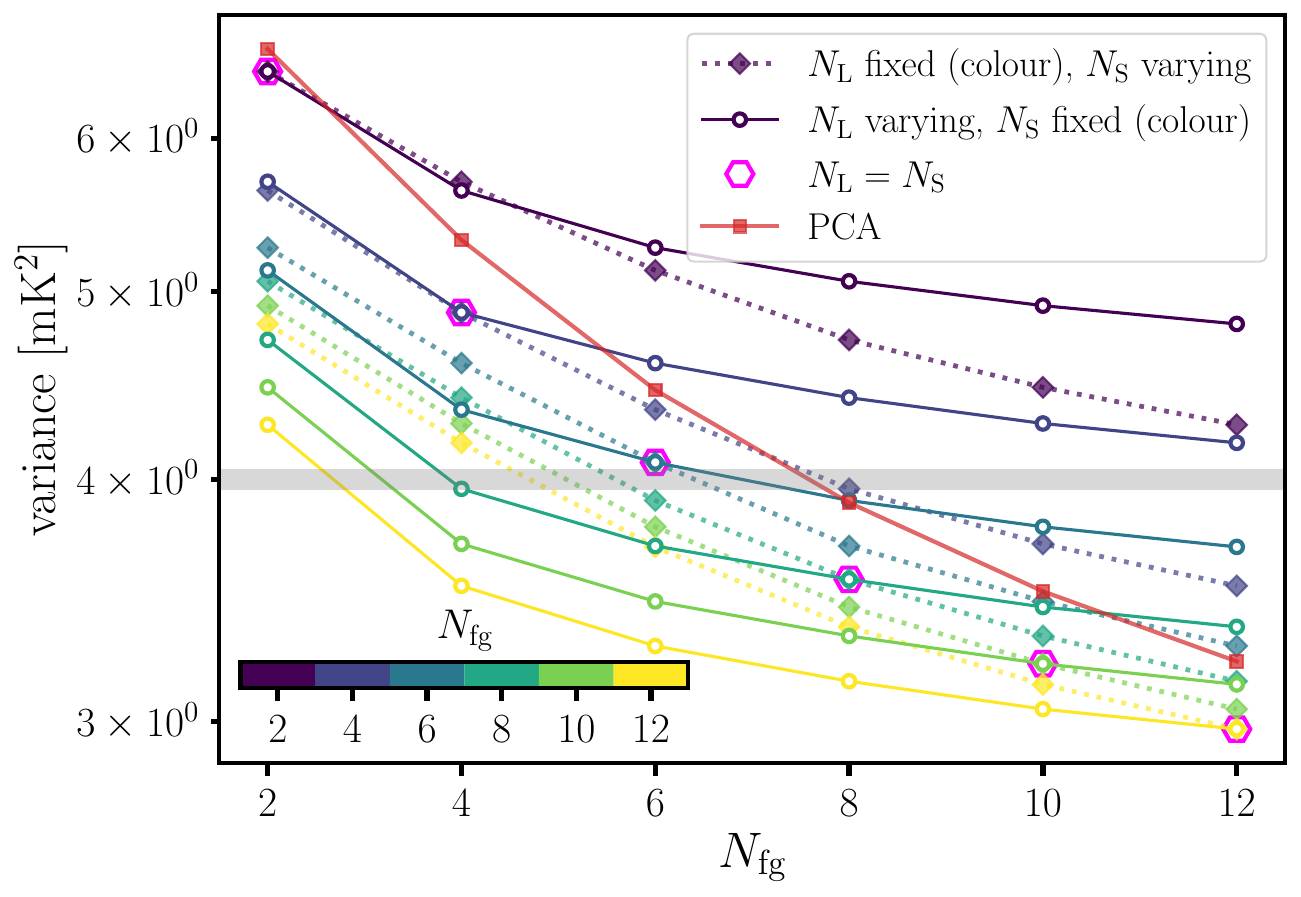}
    \caption{Variance in the mPCA cleaned cubes varying the number of components removed. With dotted lines and filled symbols, we refer to cleaning by fixing the number of the large-scale components removed, $N_{\rm L}$, given by the colour of the line (inset colour scale), and with solid lines and empty symbols, the other way around: $N_{\rm S}$ is kept fixed (colour) and $N_{\rm L}$ is varying ($x-$axis). With empty magenta symbols we highlight the cubes cleaned with $N_{\rm L}=N_{\rm S}$, for a more direct comparison with the PCA results, shown by the red squares and solid line. We show the estimated uncorrelated noise level with a grey band. }
    \label{fig:variance_mPCA_NLNS}
\end{figure}

We look at the variance of the cleaned mPCA cubes as a function of the number of components removed. Although we decided to work with the CF footprint only, in Fig.~\ref{fig:variance_mSVD} we show the variance of the OF cubes cleaned with mPCA, to check the effects of cropping and of weighting for mPCA (and mSVD) too. As in Fig.~\ref{fig:variance_foot}, the solid line in blue (green) corresponds to an mPCA cleaning applied to the OF (CF), to be compared with the weighted counterpart in dotted red (brown) and to the effect of cropping the OF cleaned results to the CF (solid orange of mPCA and dotted violet for weighted-mPCA); pink (grey) bands are the expected noise floor for the OF (CF) observed cubes. On the $x$-axis we show the number of modes removed, fixing $N_{\rm L}=N_{\rm S}$. The conclusions that we can draw are similar (but stronger) to those we drew from Fig.~\ref{fig:variance_foot}: (i) weighting alone is not sufficient for dealing with the high-variance pixels, 
and (ii) the cropping is indeed beneficial for the mPCA solution to be more descriptive hence efficient for our data cube.
We do not show any other weighted mPCAw (or mSVD, mSVDw) results, as we do not find appreciable differences with respect to mPCA. 

It is informative to compare Figs.~\ref{fig:variance_foot} and~\ref{fig:variance_mSVD}: our first direct comparison of the multiscale cleaning versus the standard PCA. Even when working within the OF, the mPCA residual cube reaches the noise floor (in pink) with only $N_{\rm fg} =12-13$ components removed. These results do not take full advantage of the multiscale method that entails two independent subtractions, since we are imposing  $N_{\rm L}=N_{\rm S}$.

We look at the effect of varying $N_{\rm L}$ and $N_{\rm S}$ independently in Fig.~\ref{fig:variance_mPCA_NLNS}, for the CF only. With the dotted lines, we let $N_{\rm S}$ vary and keep $N_{\rm L}$ fixed, given by the colour of the line (inset colour scale); it is the other way around for the solid lines ($N_{\rm L}$ varying, $N_{\rm S}$ fixed). Increasing $N_{\rm L}$ is the primary driver of a variance drop for $N_{\rm L} \leq 4$; after four modes removed on the large scale, the variance decrease slows down (keeping $N_{\rm S}$ fixed). Conversely, the contribution of the increase of $N_{\rm S}$ to reduce the variance is more constant, also beyond the range of mode numbers explored in Fig.~\ref{fig:variance_mPCA_NLNS}. 
With the empty magenta symbols, we highlight the cases where the cleaning has been performed with $N_{\rm L}=N_{\rm S}$ for a quicker comparison with the PCA results (solid red line). We demonstrate once again that applying PCA separately on the spatial scales (i.e. mPCA) is not equivalent to applying it on the whole-scale original cube, even when the number of components removed is the same.

The results shown so far have been primarily qualitative, looking at the properties of the PCA and mPCA decompositions. In Sect.\,\ref{sec:discussion}, we discuss how demonstrative these tests are and what we have learnt from them. In the next section, we  present more quantitative results and evaluate the cleaned cubes by measuring their cross-correlation signal with the WiggleZ galaxies.

\section{Cross-power spectrum: Estimation and theoretical modelling}
\label{sec:model}

We analysed the cosmological signal present in the maps through its cross-power spectrum with the position of the 4031 WiggleZ galaxies, $P_{\hi,\text{g}}$. For the estimator and theoretical modelling, we closely followed CL23. 

\subsection{Cross-power spectrum estimation}
\label{sec:Pkest}

The MeerKAT data cube is in celestial coordinates and receiver channel frequency (R.A.,
Dec., $\nu$). To perform fast Fourier transform (FFT) calculations on data, we re-grid 
the cube onto regular three-dimensional Cartesian coordinates $\mathbf{x}$ with lengths $l_{\text{x}_1},\,l_{\text{x}_2},\,l_{\text{x}_3}$ in $h^{-1}$Mpc, assuming a \textit{Planck18} cosmology \citep{Planck:2018nkj} and voxel volume, $V_{\rm cell}$.

The Fourier transforms of the \hi~temperature maps $\delta T_\hi$ and the galaxy number field $n_\text{g}$ are defined as
\begin{equation}
    \tilde{F}_\hi(\mathbf{k})=\sum_{\mathbf{x}} \delta T_\hi(\mathbf{x}) w_\hi(\mathbf{x}) \exp (i \mathbf{k}{\cdot}\mathbf{x})\, ,
\end{equation}
\begin{equation}
    \tilde{F}_\text{g}(\mathbf{k})=\sum_{\mathbf{x}} n_\text{g}(\mathbf{x}) w_\text{g}(\mathbf{x}) \exp (i \mathbf{k}{\cdot}\mathbf{x}) - N_\text{g}\tilde{W}_\text{g}(\mathbf{k})\,,
\end{equation}
where $N_\text{g}\,{=}\,\sum n_\text{g}$ is the total number of galaxies, $\tilde{W}_\text{g}$ is the weighted Fourier transform of the normalised selection function, $W_\text{g}$, which takes care of the incompleteness of the galaxy survey, as designed by the WiggleZ collaboration \citep{Blake:2010xz}, and the weights $ w_\hi(\mathbf{x})$ and $ w_\text{g}(\mathbf{x})$ are the inverse variance map introduced in Sect.\,\ref{sec:data} and the optimal weighting of \cite{Feldman:1993ky}, for the \hi~and galaxy fields respectively.

Additionally, to limit the ringing in the spectra calculations (i.e. spurious correlations between adjacent $k$-bins), we apply tapering functions that smoothly suppress the cleaned data to zero at the edges.
CL23 apodised the cube in the frequency direction using a Blackman window \citep{blackman1958}; in the angular direction, the weight maps $w_\hi(\mathbf{x})$ fall off at the edges due to scan coverage and effectively act as tapering, hence no further tapering was needed. In this analysis, we keep the Blackman choice in the line-of-sight direction, but since we use a smaller footprint---the CF---the weights within the CF no longer act as tapering. Instead, we adopt a Tukey window function\footnote{The Blackman window function is more optimal and designed to have close to the minimal leakage possible. However, in the angular direction, we cannot afford it with the present number of pixels; we will improve on this with future observations with a larger area coverage.} to maximise the number of pixels whose intensity information is retained in the power spectrum computation \citep{windows4FT}; we display it in the left panel of Fig.~\ref{fig:foot} with the yellow curves---dotted and dashed for when the window reaches 0.5 and 0, respectively.

Finally, we define the cross-power spectrum estimator as
\begin{equation} \label{eq:Pkcross}
    \hat{P}_{\hi,\text{g}}(\mathbf{k}) = \frac{V_\text{cell}}{\sum\limits_\mathbf{x} w_\hi(\mathbf{x})w_\text{g}(\mathbf{x})W_\text{g}(\mathbf{x})}\operatorname{Re}\left\{\tilde{F}_\hi(\mathbf{k}){\cdot} \tilde{F}^{*}_\text{g}(\mathbf{k})\right\}\frac{1}{N_\text{g}}\,.
\end{equation}
All spectra are averaged into bandpowers $\lvert\mathbf{k}\rvert\,{\equiv}\,k$ yielding the final spherically averaged power spectra measurements that we compare against theory.

Our cleaned intensity maps are highly noise-dominated 
(see Sect.\,\ref{sec:data}, i.e. the noise level is of about $2-3$ mK), hence we use Gaussian statistics to analytically estimate the uncertainties of the cross-spectrum measurements:
\begin{equation}\label{eq:Pkerr}
    \hat{\sigma}_{\hi,\text{g}}(k)=\frac{1}{\sqrt{2 N_\text{m}(k)}} \sqrt{\hat{P}^2_{\hi,\text{g}}(k)+\hat{P}_\hi(k)\left(\hat{P}_\text{g}(k)+\frac{1}{\bar{n}_\text{g}}\right)}\,,
\end{equation}
where $N_\text{m}$ is the number of modes in each $k$-bin and $\bar{n}_\text{g}\,{=}\,N_\text{g}/(l_{\text{x}_1}\,{\times}\,l_{\text{x}_2}\,{\times}\,l_{\text{x}_3})$ is the number density of galaxies. The auto-spectra of the fields, $\hat{P}_\hi$ and $\hat{P}_{\text{g}}$, are computed in the same fashion of Eq. (\ref{eq:Pkcross}), namely,
\begin{equation}\label{eq:HIautoPk}
    \hat{P}_\hi(\mathbf{k}) = \frac{V_\text{cell}}{\sum_\mathbf{x} w_\hi^2(\mathbf{x})}\lvert\tilde{F}_\hi(\mathbf{k})\rvert^2\,,
\end{equation}
\begin{equation}
    \hat{P}_\text{g}(k) = \frac{V_\text{cell}}{\sum_\mathbf{x} w^2_\text{g}(\mathbf{x})W^2_\text{g}(\mathbf{x})} \left[|\tilde{F}_\text{g}(\mathbf{k})|^2 - N_\text{g}\sum_\mathbf{x} w_\text{g}^2(\mathbf{x})W_\text{g}(\mathbf{x})\right]\frac{1}{N^2_\text{g}}\,,
\end{equation}
where we included a shot noise term in the galaxy auto-spectrum, whereas $\hat{P}_\hi$ includes (and is dominated by) thermal noise. CL23 compared these analytical error estimations to the ones calculated from cross-correlating the MeerKAT data with the random WiggleZ catalogues used to derive the
selection function, finding good agreement. In Sect.\,\ref{sec:crossPk}, we further test the Gaussian error assumption by means of a reduced $\chi^2$ analysis and comparing them directly with the data-driven jackknife errors, finding also good agreement.

\begin{figure}
\centering
    \includegraphics[width=0.9\columnwidth]{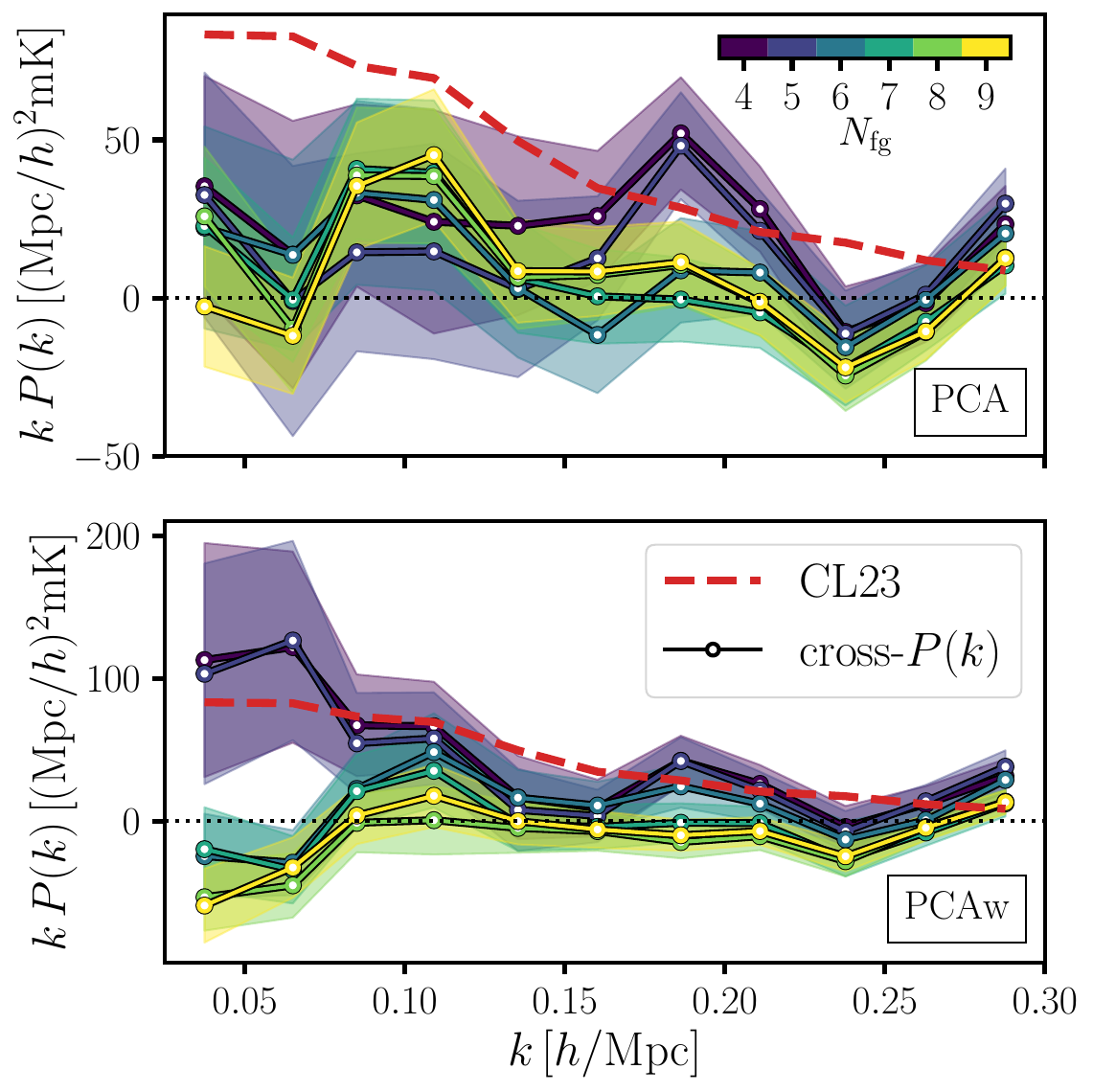}
    \caption{Cross-power spectra between WiggleZ galaxies and MeerKAT \hi~intensity maps at redshift $z\approx0.4$ for the PCA and PCAw cleaned cubes in the top and bottom panel, respectively. Maps are cleaned by removing different numbers of PCA (and PCAw) modes, from dark, $N_{\rm fg} = 4$, to light colour lines, $N_{\rm fg} = 9$; $1\sigma$ error bars are shown with the corresponding shaded areas. The red dashed line is the fitted model from CL23 (same data, different cleaning strategy) that instead includes a $\cal{T}$-correction for the signal loss; as a rule-of-thumb, the uncorrected CL23 best-fit would show an $80\%$ lower amplitude at the largest scale probed (e.g. below the $1\sigma$ boundary of the $N_{\rm fg} = 4$ and 5 PCAw measurements for $k\lesssim 0.12\, h/$Mpc in the bottom panel). We plot the product $k P(k)$ to accentuate the small scales that would otherwise be flattened by the much higher low-$k$ measurements.
    } 
    \label{fig:crossPK}
\end{figure}

\begin{figure}
\centering
    \includegraphics[width=0.9\columnwidth]{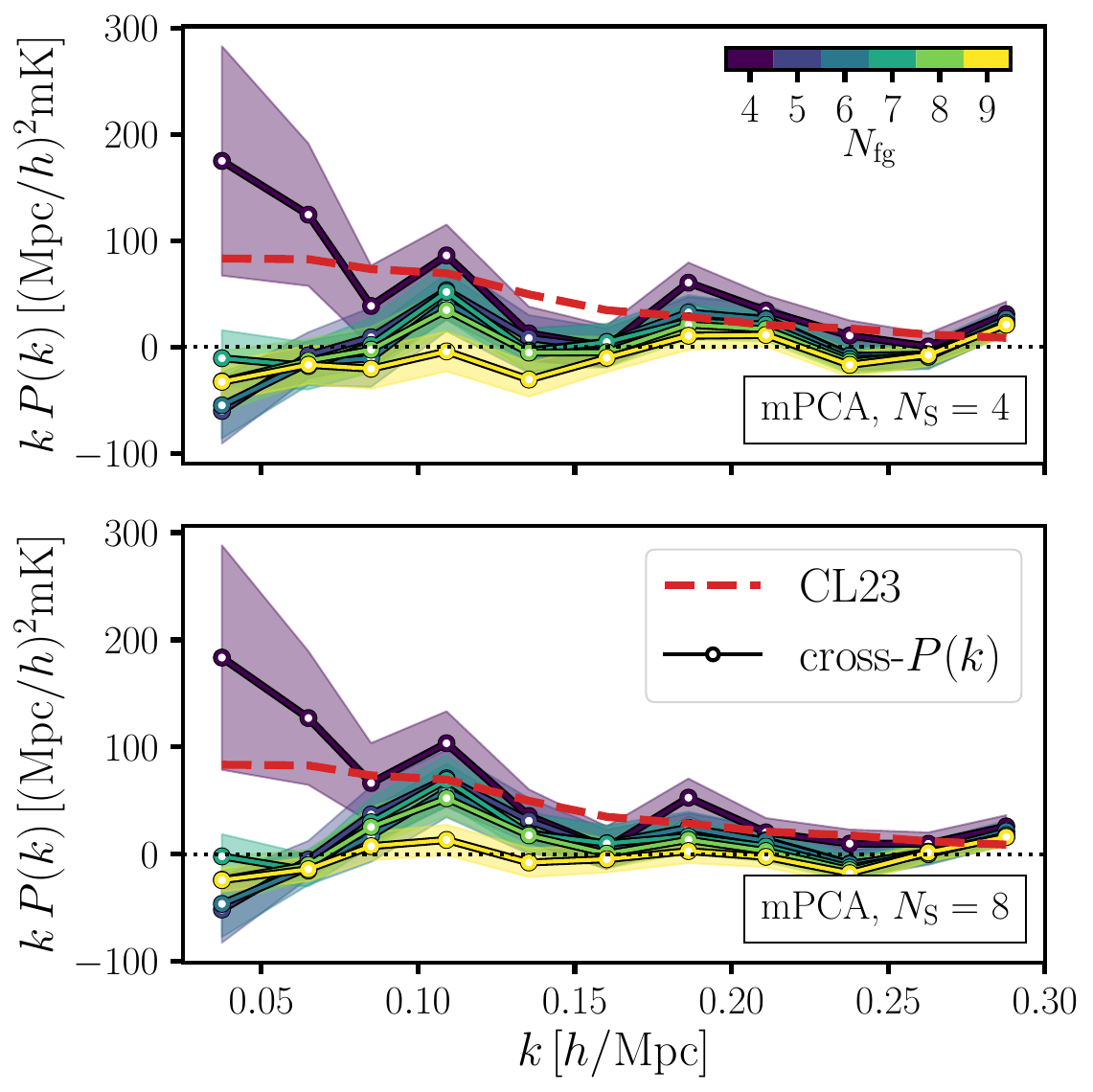}
    \caption{Cross-power spectra between WiggleZ galaxies and MeerKAT \hi~intensity maps cleaned by mPCA at redshift $z\approx0.4$. We keep the number of modes removed at small scale fixed at $N_{\rm S} = 4$ (top panel) and $N_{\rm S} = 8$ (bottom),  and we remove different numbers of large-scale modes (from dark, $N_{\rm L} = 4$, to light colour lines, $N_{\rm L} = 9$) , with $1\sigma$ error bars (corresponding shaded areas). The red dashed line is the fitted model from CL23 (same data, different cleaning strategy) that instead includes a $\cal{T}$ correction for the signal loss; as a rule-of-thumb, the uncorrected CL23 best-fit would show an $80\%$ lower amplitude at the largest scale probed (e.g. below the $1\sigma$ boundary of the $N_{\rm L} = 4,$ measurements for $k\lesssim 0.14\, h/$Mpc). We plot the product  $k P(k)$ to accentuate the small scales that would otherwise be flattened by the much higher low-$k$ measurements. 
   } 
    \label{fig:crossPK_mPCA}
\end{figure}

\subsection{Theoretical modelling}\label{sec:Model}

Our fiducial cosmological model is \textit{Planck18}. We assume it to compute the linear matter power spectrum $P_\text{m}$, derived through \texttt{CAMB} \citep{Lewis:1999bs}, and all other cosmological quantities (see below). We work at the effective redshift $z_{\rm eff} = 0.42$, corresponding to the median frequency of the data cube.

We adopt the following cross-power spectrum model:
\begin{multline}\label{eq:Pkmodel}
    P_{\hi,\text{g}}(\mathbf{k}) = \overline{T}_\hi b_\hi b_\text{g} r (1+f\mu^2)^2 \,P_\text{m}(k) \\ \times\, \exp\left[\frac{-(1-\mu^2)k^2 R_\text{beam}^2}{2}\right]\,.
\end{multline}
$\overline{T}_\hi$ is the mean \hi\ temperature of the field that, at fixed cosmology, depends on the overall neutral hydrogen cosmic abundance $\Omega_{\hi}$ \citep{Furlanetto2006}:
\begin{equation}\label{eq:TbarModelEq}
    \overline{T}_\hi(z) = 180\,\Omega_{\hi}(z)\,h\,\frac{(1+z)^2}{\sqrt{\Omega_\text{m}(1+z)^3 + \Omega_\Lambda}} \, {\text{mK}} \,,
\end{equation}
where $\Omega_\text{m}$ and $\Omega_\Lambda$ are the density fractions for matter and the cosmological constant, respectively. 
$b_\hi$ and $b_\text{g}$ are the \hi\ and galaxy large-scale linear biases and $r$ is their cross-correlation coefficient. We account for linear matter redshift-space distortions (RSD) with the $(1+f\mu^2)^2$ factor \citep{Kaiser:1987qv}, where $f$ is the growth rate of structure and $\mu$ is the cosine of the angle from the line-of-sight. The Gaussian damping in the last term approximates the smoothing effect of the telescope beam; in particular, we compute the corresponding standard deviation \citep{Matshawule:2020fjz}
\begin{equation}\label{eq:sigbeamsize}
    \sigma_\text{beam}(\nu) = \frac{\theta_{\rm FWHM}(\nu)}{2\sqrt{2\ln 2}}\,,
\end{equation}
through the angular resolution FWHM parameter $\theta_{\rm FWHM}$ of Eq. (\ref{eq:thetaFWHM}),
and multiply it by the comoving angular diameter distance to the effective redshift---median frequency---of the observed data set, yielding $R_\text{beam}\,{=}\,10.8\,\text{Mpc}\,h^{-1}$. Given the current noise in the data and uncertainties of the measurements, we find the above modelling of beam smearing to be sufficient for this analysis. 

For a direct comparison, we sample the power spectrum template in Eq. (\ref{eq:Pkmodel}) onto the same Fourier-space grid as the data, and convolve it with the same window functions. Together with fixing cosmology, we  
set the galaxy bias $b_\text{g} = \sqrt{0.83}$ \citep{Blake2011}. The only free parameter left to be fitted against data is the combination $\Omega_\hi b_\hi r$, namely, the amplitude of the cross-power spectrum of Eq. (\ref{eq:Pkmodel}). 

For reference, CL23 measured $\Omega_\hi b_\hi r = (0.86 \pm 0.10)\times10^{-3}$, derived using a transfer function ($\cal{T}$) approach to compensate for the  signal loss, namely, enhancing the measurements by about $80\%$ for $k< 0.15 h/$Mpc down to $20\%$ at the smallest scale considered. We did not perform the $\cal{T}$-correction on the spectra measured in this analysis, for reasons that will become clear in later sections, yet delivering a cross-spectra in agreement (and even with higher amplitude) with CL23.

\section{Capitalising on our benchmark: Measuring the correlation with the WiggleZ galaxies}
\label{sec:crossPk}

We assessed the new cleaning pipeline choices and methods described in this work, confronting the resulting cleaned cubes with the cross-correlation measurement already reported in CL23. To focus on the cleaning strategy's influence on the results, here we use the same tools presented in CL23 for computing and modelling $P_{\hi,\text{g}}(k)$, fitting its amplitude, and evaluating the goodness of the fit. The only exception to the previous statement is related to using a 2D window function before the Fourier transform that replaces the effective tapering due to the weight map of the OF (see Sect.\,\ref{sec:model}).

In many figures in this section, we report the CL23 result as a reference. In particular, we show the fitted model of Eq. (\ref{eq:Pkmodel}) with the amplitude derived by CL23---not the actual $P_{\hi,\text{g}}(k)$ data points---for clarity and to overcome the fact that in CL23 maps had been smoothed, effectively reducing their native resolutions that instead we keep here. However, the reader has to keep in mind that CL23 derived the final power spectrum with the transfer function , $\cal{T}$, procedure, whereas we stress that none of the results of the present work have a $\cal{T}$-correction applied. It might seem an unfair comparison, but it will make our point: we have improved the cleaning procedure so that the cosmological signal loss is more under control.

\subsection{The cross-power spectrum}

We start with the PCA-like methods' cleaned cubes. We show the cross-$P(k)$  in Fig.~\ref{fig:crossPK} in the top for a simple PCA analysis, in the bottom for its weighted version PCAw. Both SVD results, weighted and not, are in between those presented; that is to say: PCA and PCAw bracket the range of the PCA-like analyses. In each panel, we show the cross-$P(k)$ for different cubes cleaned with different numbers $N_{\rm fg}$ of removed components, colour coded from dark ($N_{\rm fg} = 4$) to light ($N_{\rm fg} = 9$). The lines and empty symbols are the measurements, and the corresponding shaded areas are their $1\sigma$ uncertainties. Besides the encouraging agreement of some of these results with CL23 (red dashed line), the overall behaviour of these power spectra is promising (especially for PCAw) as our interpretation follows. We do not show any $N_{\rm fg} < 4$ results as they are widely oscillating, namely, too much foreground contamination is still in the maps, and no measurement is possible. With the $N_{\rm fg} = 4$ cleaned cube, the $P_{\hi,\text{g}}(k)$ eventually reaches a reasonable `shape'; indeed, besides the amplitude, the consistency with CL23 indicates, first of all, a theoretically expected $P(k)$ behaviour. As contaminant suppression becomes more aggressive (increasing $N_{\rm fg}$) the cosmological signal is increasingly suppressed as well, and the $P_{\hi,\text{g}}(k)$ gradually loses power. Eventually, for $N_{\rm fg} = 6-7$, we are fully compatible with a null spectrum: the correlation is lost.

Next, we computed the cross-$P(k)$ of the mPCA-cleaned cubes. The results are shown in Fig.~\ref{fig:crossPK_mPCA} where, with the same colour codes of Fig.~\ref{fig:crossPK}, we show different cleaning with dark to light colour lines and areas (mean values and $1\sigma$ errors) and the red dashed line is the CL23 fitted model. For the mPCA case, we can choose how many modes to remove on the large ($N_{\rm L}$) and---independently---on the small scale ($N_{\rm S}$); the spectra in the figure have been derived varying $N_{\rm L}$ and keeping $N_{\rm S}$ fixed to 4 (top panel) and 8 (bottom). 
We find a similar---and actually clearer---trend as what we described above for Fig.~\ref{fig:crossPK}: the cleaned cube that gives the highest $P_{\hi,\text{g}}(k)$ amplitude corresponds to a `sweet spot' when varying the $N_{\rm fg}$ parameters. For brevity, we decided not to show results for varying $N_{\rm S}$; we find that 
the large scale $N_{\rm L}$ is the primary driver of the overall amplitude of the cross-$P(k)$.

\subsection{Least-squares fitting}

\begin{figure}
\centering
    \includegraphics[width=0.9\columnwidth]{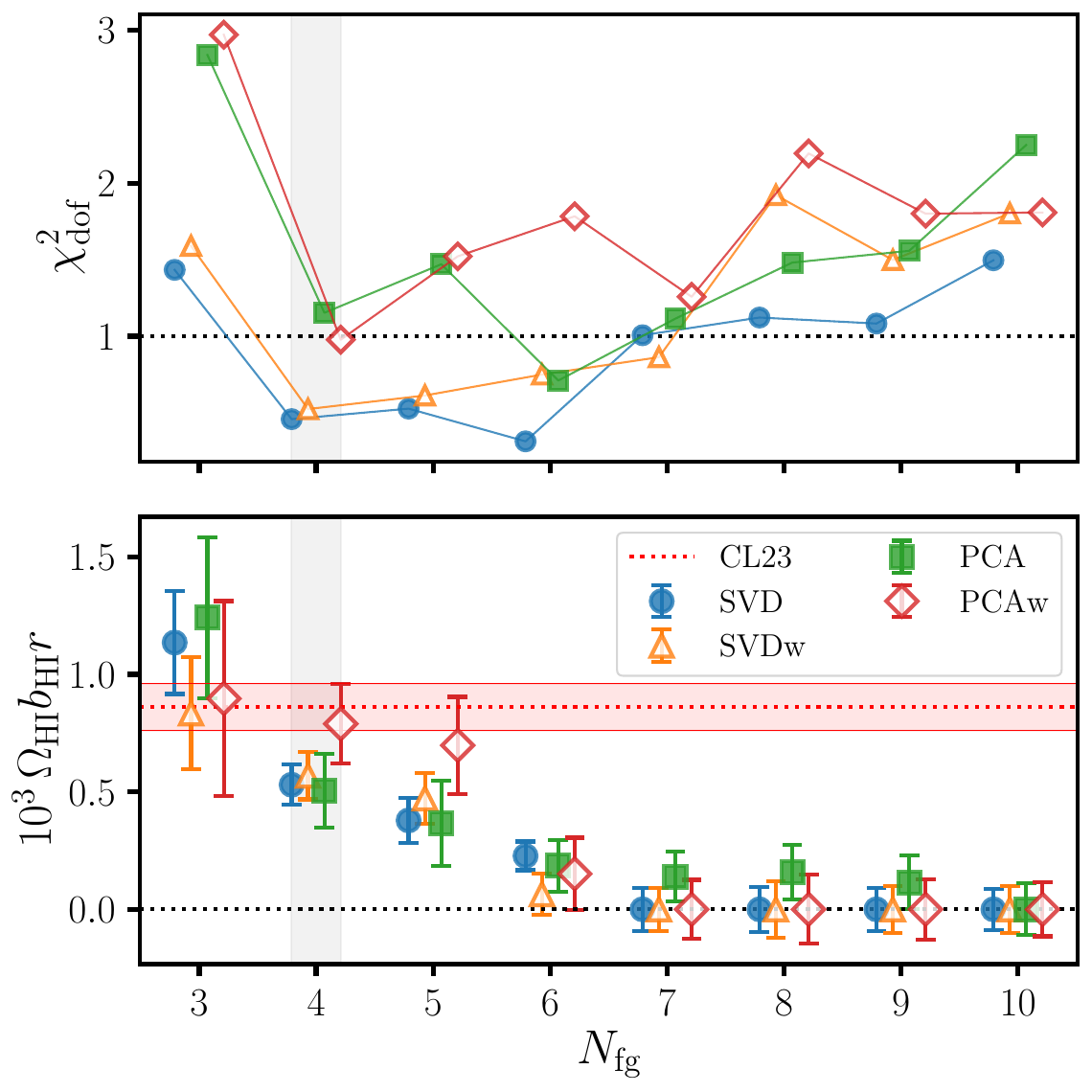}
    \caption{Sensitivity of our results to the PCA-like contaminant subtractions as function of how many modes have been removed from the original cube. In the top panel, we show the reduced $\chi^2$ for each foreground cleaned cross-power spectra relative to its best-fitting model; in the bottom panel, the corresponding best-fitting $\Omega_{\hi} b_{\hi} r$ values and their $1\sigma$ error bars. Filled symbols, circles and squares, are for an SVD a PCA cleaning respectively; their weighted counterparts corresponds to empty triangles and diamonds. In the bottom panel, the shaded red region highlights the CL23 best-fitting value and $1\sigma$ uncertainty that was obtained with a PCAw $N_{\rm fg}=30$ cleaning and a $\cal{T}$-correction (the `uncorrected' CL23 amplitude would lay between $0.2-0.3 \times 10^{-3}$). We highlight in grey the $N_{\rm fg}=4$ results that correspond to our best and reference PCA result: the smallest $N_{\rm fg}$ to reach a plausible $\chi^2$.} 
    \label{fig:fit_results}
\end{figure}

\begin{figure}
\centering
    \includegraphics[width=0.9\columnwidth]{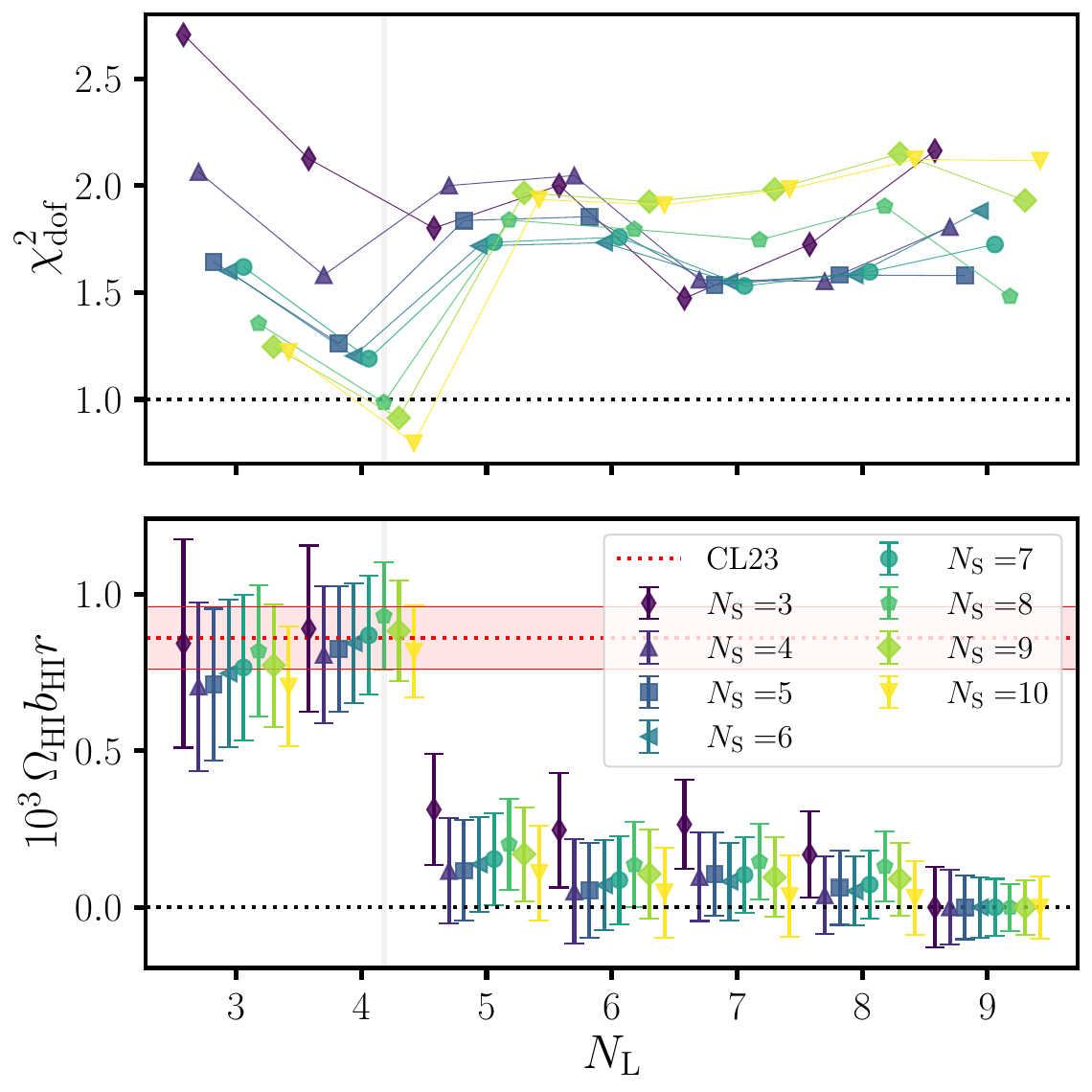}
    \caption{Sensitivity of our results to the mPCA contaminant subtraction as function of how many modes have been removed from the original cube. The top panel shows the reduced $\chi^2$ for each foreground cleaned cross-power spectra relative to its best-fitting model. The bottom panel the corresponding best-fitting $\Omega_{\hi} b_{\hi} r$ values and their $1\sigma$ error bars.  
    All quantities are plotted as a function of the $N_{\rm fg}$ pair of the mPCA analysis, with $N_{\rm L}$ on the $x$-axis and $N_{\rm S}$ colour- and symbol-coded spanning from $N_{\rm S}=3$ in violet to $N_{\rm S}=10$ in yellow. For comparison, the shaded red region highlights the CL23 best-fitting value and $1\sigma$ uncertainty, that was obtained with a $\cal{T}$-correction (the `uncorrected' CL23 amplitude would lay between $0.2-0.3 \times 10^{-3}$). We highlight in grey the $N_{\rm L}, N_{\rm S}=4, 8$ results that correspond to our best and reference mPCA result. Results from mSVD and the weighted mSVDw and mPCAw show negligible differences.} 
    \label{fig:fit_results_mPCA}
\end{figure}

Next, we conducted a least-squares fit to the measured $P_{\hi,\text{g}}(k)$ with the model described in Sect.\,\ref{sec:model}, and measure their amplitude  
$\Omega_{\hi} b_{\hi} r$.  
We assumed a Gaussian covariance, effectively ignoring correlation among adjacent $k$-bins\footnote{In ongoing work with higher signal-to-noise MeerKAT \hi~intensity mapping data, we are going beyond the Gaussian covariance approximation, e.g. \citet{MK_GAMA}.}. We fit for one parameter, the amplitude of the cross-$P(k)$, and worked with 11 $k$-bins (data points), yielding 10 degrees of freedom. 
The results are in Fig.~\ref{fig:fit_results} for the PCA cubes and Fig.~\ref{fig:fit_results_mPCA} for mPCA as a function of the number $N_{\rm fg}$ of modes removed.

We start by looking at the PCA-framework results in Fig.~\ref{fig:fit_results}. The top panel shows the reduced $\chi^2_{\rm dof}$ of the best-fitting model; 
the bottom panel shows the fitted amplitudes with their $1\sigma$ uncertainties and, with the red dotted line and shaded area, the CL23 result and $1\sigma$ uncertainty. Different symbols and lines correspond to differently cleaned cubes: SVD (blue circles), PCA (green squares) and their weighted counterpart (empty triangles and diamonds for SVDw and PCAw respectively). 
Among those $N_{\rm fg}$ solutions in good agreement with the model ($\chi^2_{\rm dof} \approx 1$), $N_{\rm fg} = 4$ is the lowest $N_{\rm fg}$ and corresponds to the highest significance (highest amplitude) result, independently of the cleaning pipelines shown. The story told by Fig.~\ref{fig:fit_results} quantitatively matches our interpretation of the power spectra shown in Fig.~\ref{fig:crossPK}: the power spectra are unrealistic for $N_{\rm fg}<4$, for $N_{\rm fg}\geq 4$ the theoretical model can eventually describe them and we progressively see suppression of the power (signal) for $N_{\rm fg}>4$, finally reaching a null detection for $N_{\rm fg}>7-8$. Our best result---the best compromise between suppression of contaminants and suppression of the cosmological signal---is given by a PCAw analysis with $N_{\rm fg}=4$. We notice that PCA and PCAw give the lowest and highest amplitude among the four cleaned cubes. Nevertheless, all four of them are in agreement: they are less than $1\sigma$ away ($2\sigma$ for PCA) the $\cal{T}$-corrected CL23 amplitude, and higher than the uncorrected CL23 measured amplitude ($\approx0.25 \times 10^{-3}$), suggesting we experience less signal loss in this analysis compared to CL23.  
Another point made by this plot is about weighting that improves the PCA and SVD solutions. Indeed, PCAw and SVDw (i) give more reasonable fits than PCA and SVD ($\chi^2_{\rm dof}$ closer to 1, top panel), (ii) exhibit more clearly the `sweet' spot at $N_{\rm fg}=4$---it is unambiguously the minimum of their $\chi^2_{\rm dof}$ curves--- and, for those good fits, (iii) correspond to the highest amplitude of the cross-spectra. We recall here  that although both signal loss and residual contaminants are expected in the cleaned cubes, we are not concerned about the latter in the cross-correlation, as those remaining contaminants should not correlate with the galaxy positions; hence, we can consider these measurements as lower limits and, thus, the highest amplitude is closest to the true one.

We now focus on the least-square fitting of the mPCA spectra; results are summarised in Fig.~\ref{fig:fit_results_mPCA}. The top panel shows the reduced $\chi^2_{\rm dof}$ of the fits and the bottom the derived amplitude values and their standard deviations. On the $x$-scale, we show the number of coarse-scale removed modes $N_{\rm L}$; instead, different values of $N_{\rm S}$ correspond to different colours and symbols.  
We can confirm it is the $N_{\rm L}$ parameter that mainly drives the overall amplitude; however, we need to set also $N_{\rm S}$ accordingly to yield the best fits, in terms of the $\chi^2$ value. For example, all $N_{\rm L} = 4$ amplitude values lie in the expected range (CL23 result); yet, we need $N_{\rm S}>5$ to get a satisfactory $\chi^2$, with, for instance, $N_{\rm S}=3$ always leading to an arguably high $\chi^2$ regardless of the $N_{\rm L}$ we set. We can spot another interesting trend: the $N_{\rm S}=8$ cases maximise the amplitude for any given $N_{\rm L}$. We can speculate that here $N_{\rm L}$ plays the most important role because we are fitting for an amplitude rather than a shape, and indeed $N_{\rm S}$ is important for reaching a good  $\chi^2$. Hence, the small-scale cleaning will become crucial when data quality will allow us to constrain cosmology, namely, the shape of the power spectrum. 

As we did for the PCA-like methods, among the cubes with an acceptable $\chi^2$, as the mPCA reference, we selected  the cube with the highest amplitude (and, hence a higher detection significance), which is given by the $N_{\rm L}, N_{\rm S}=4,8$ combination.

We stress again that, for all the cleaning strategies reported in Figs.~\ref{fig:fit_results} and~\ref{fig:fit_results_mPCA}, the best-performing residuals (good fit and highest amplitude) correspond to values of $\Omega_{\hi} b_{\hi} r$ that are in full agreement with the $\cal{T}$-corrected CL23 result. Yet, their statistical uncertainties are higher than the $12\%$ for CL23: $30\%$ for PCA, $21\%$ for PCAw and $18\%$ for mPCA. 
However, CL23 (i) performed the extra `reconvolution' smoothing, suppressing power, and (ii) removed more components (30 versus 4) suppressing thermal noise\footnote{Uncorrelated noise suppression resembles signal loss; both are filtered similarly due to their similar frequency structure.}; both decrease the cubes auto-spectra, influencing the Gaussian errors' calculations.
Regardless, CL23 experienced a degeneracy with the number of removed modes---it was less clear which $N_{\rm fg}$ to set, leading to an additional variance of $14\%$. Interestingly, when we account for this error in the overall $\Omega_{\hi} b_{\hi} r$  uncertainty, we find it comparable with what is estimated in the present analyses.

\begin{figure}
\includegraphics[width=\columnwidth]{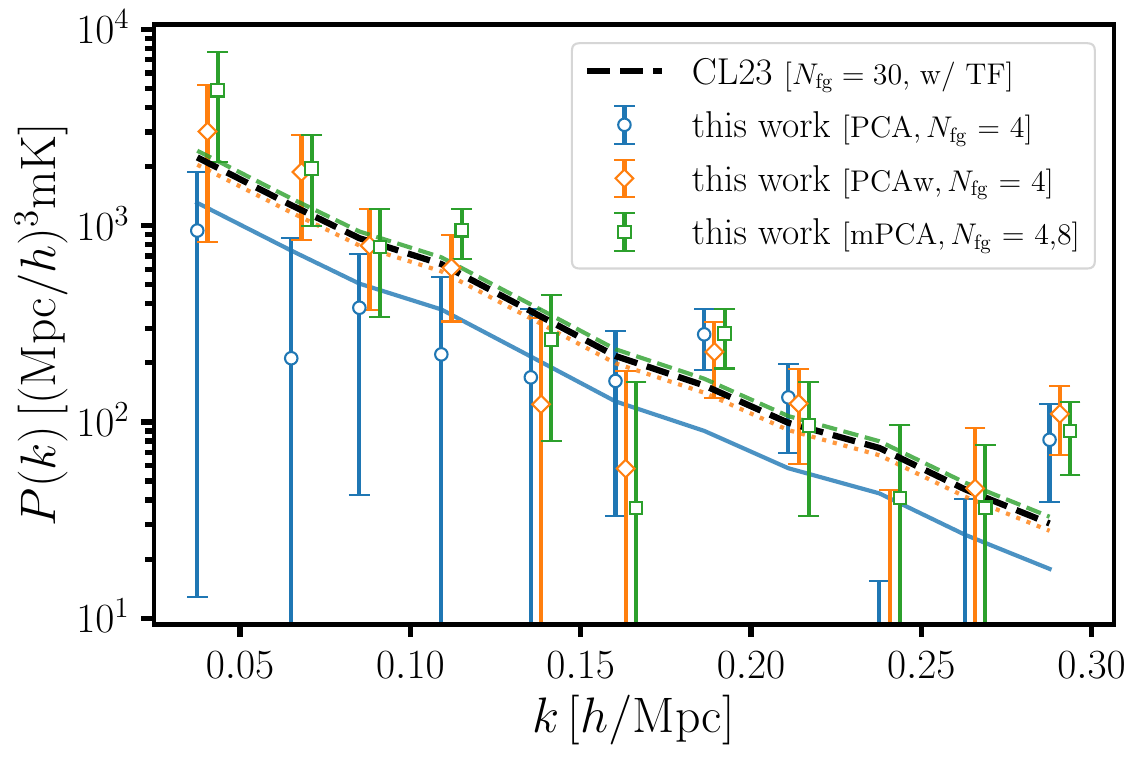}
    \caption{Cross-power spectra between WiggleZ galaxies and MeerKAT \hi~intensity maps at redshift $z\approx0.4$. Maps have been cleaned by a PCA and PCAw analysis with $N_{\rm fg}=4$ (blue circles and orange diamonds respectively) and by mPCA with $N_{\rm L},N_{\rm S}=4,8$ (green squares). We show error bars with $1\sigma$ uncertainties and the resulting best-fit curves (solid blue for PCA, dotted orange for PCAw and dashed green for mPCA).
    The black dashed line is the fitted model from CL23 (same data, different cleaning strategy) that instead was enhanced by a transfer function correction.}
    \label{fig:results}
\end{figure}

We summarise the results from all contaminant-separation pipelines in Table~\ref{table:summary},  and we further showcase the PCA, PCAw and mPCA solutions in Fig.~\ref{fig:results}, with the results from CL23 for comparison.
Table~\ref{table:summary} displays the cleaning level ($N_{\rm fg}$ or $N_{\rm L}$, $N_{\rm S}$), whether or not a transfer function correction is in place ($\cal{T}$), the measured amplitude of the cross-spectra ($\Omega_{\hi}b_{\hi}r$), the goodness of the fit ($\chi^2_{\rm dof}$), and its statistical detection significance $\sqrt{\Delta \chi^2}\,{\equiv}\,\sqrt{\chi^2_{\rm null} - \chi^2}$, namely, the difference between the $\chi^2$ evaluated using our cross-correlation model, and one using a null model with zero power. The uncertainty that we quote for the CL23 amplitude is the sum in quadrature of the two reported errors ($\Omega_{\hi} b_{\hi} r =[0.86 \pm 0.10 ({\rm stat}) \pm 0.12 ({\rm sys})]    \times 10^3$); its significance has also been updated accordingly.

\subsection{Null tests and jackknife errors}

\begin{figure*}
\centering
    \includegraphics[width=1.8\columnwidth]{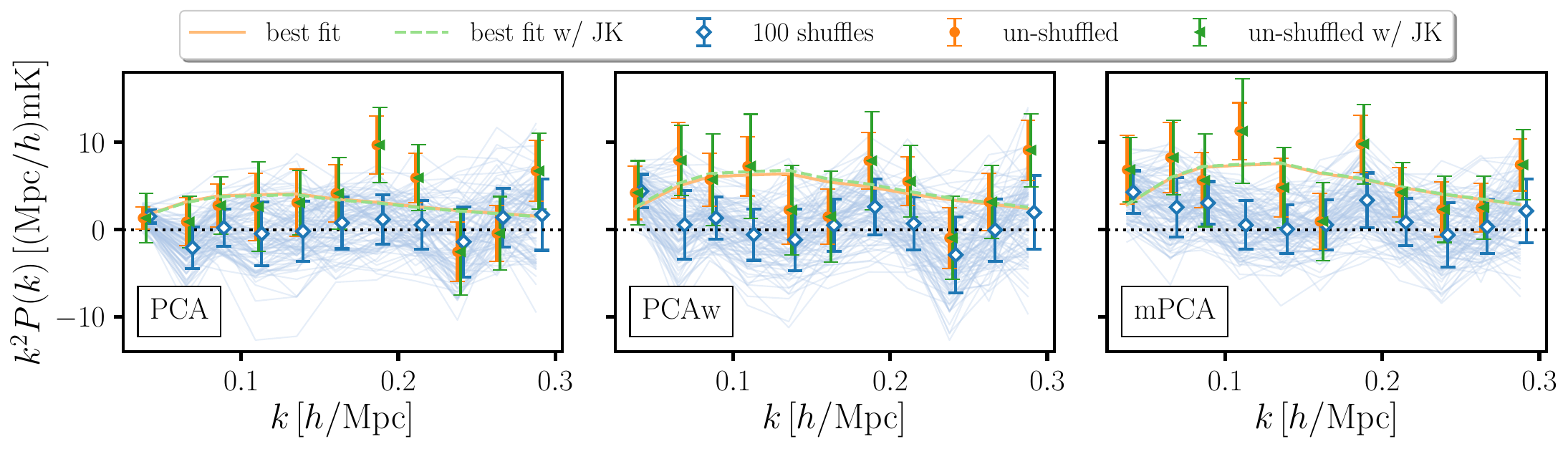}
    \caption{Null tests for the cross-power spectra derived with the PCA $N_{\rm fg}=4$ (left panel), its weighted version PCAw $N_{\rm fg}=4$ (middle) and multiscale version mPCA $N_{\rm L}, N_{\rm S}=4, 8$ (right). The WiggleZ galaxy maps have been shuffled along frequency (redshift). The thin blue lines show the spectra computed for 100 different shuffles. Empty blue diamonds correspond to the average and standard deviation across the shuffled samples, that we compare against (i) the reference spectra (filled orange circles) and (ii) the reference spectra with jackknife errors (filled green triangles). We use the solid orange and dashed green lines to show the best fits of the reference spectra, using Gaussian and jackknife errors respectively.
    } 
    \label{fig:null_t}
\end{figure*}

To validate the detections discussed above, we perform null tests by randomising galaxies along the frequency direction, see Fig.~\ref{fig:null_t}. Each panel refers to a different reference cube (from left to right: PCA $N_{\rm fg}=4$, PCAw $N_{\rm fg}=4$ and mPCA $N_{\rm L}, N_{\rm S}=4,8$), whose cross-$P(k)$ is displayed with orange dots. The 100 cross-$P(k)$ with the 100 randomised galaxy fields are in light blue, their overall mean and standard deviation correspond to the empty diamonds: fully compatible with zero.

We believe that the uncertainties of our measured $P_{\hi,\text{g}}(k)$ are adequate as the reasonable $\chi^2_{\rm dof}$ of the fitted reference $P_{\hi,\text{g}}(k)$ indicate. However, we further validate them by comparing them with jackknife errors \citep{Norberg2009}. We divide the cleaned cubes into 320 $N_{\rm jk}$ equal sub-cubes (using a higher $N_{\rm jk}$ does not produce worthwhile different results) and compute the 320 cross-spectra $P^{\rm jk}(k)$. The final jackknife spectrum is given by the mean $\mu_{\rm jk}(k)$ of the $N_{\rm jk}$ spectra, and its uncertainties are computed as follows \citep{Hartlap2007,Favole2021}:
\begin{equation}
    \sigma_{\rm jk}^2(k) = \frac{N_{\rm jk} - 1}{N_{\rm jk}} \frac{N_{\rm jk} - 1}{N_{\rm jk} - N_k -2} \sum_{i=1}^{N_{\rm jk}} \left[P^{\rm jk}_i(k)-\mu_{\rm jk}(k)\right]^2\,,
\end{equation}
with $N_k=11$ the number of $k$-bins.  We plot the $\{\mu_{\rm jk}(k), \sigma_{\rm jk}(k)\}$ data points in the panels of Fig.~\ref{fig:null_t} with filled green triangles to compare them with our reference spectra in orange. As expected, jackknife errors tend to be higher because of the non-independence of the jackknife sub-cubes. However, overall we find a satisfying agreement with the Gaussian errors of our reference analysis. Indeed, the best-fit of $\{\mu_{\rm jk}(k), \sigma_{\rm jk}(k)\}$ is almost indistinguishable from the reference one (dashed green versus solid orange lines), with the---slight---exception of PCAw (middle panel). Being more quantitative, compared to the reference analyses, the higher $\sigma_{\rm jk}^2(k)$ reduce the $\chi^2_{\rm dof}$ by $\sim40\%$ for all three reference cubes considered, and lead to practically unvaried amplitudes and associated error values, with the exception of PCAw: $\Omega_{\hi} b_{\hi} r \times10^3 = 0.84 \pm 0.19$ compared to the reference $0.79 \pm 0.17$.


\begin{table}
\caption{\label{table:summary}Results of the least-squares fits of the cross-power spectra of the MeerKAT \hi~IM maps and the WiggleZ galaxy at redshift $z\approx0.4$.}
\centering
\begin{tabular}{lccccc}
\hline\hline
 Method & $N_{\rm fg}$  &$\cal{T}$ & $\Omega_{\hi} b_{\hi} r \times 10^3$ & $\chi^2_{\rm dof}$ & significance\\
\hline
SVD & 4& no  & $0.63 \pm 0.13$ &0.78 & 3.9$\sigma$\\
SVDw & 4& no  & $0.75 \pm 0.16$ &0.91 & 4.4$\sigma$\\
{\bf PCA} & {\bf 4} &{\bf no}  & $\mathbf{0.50 \pm 0.16}$ &{\bf 1.15} & $\mathbf{3.3\sigma}$\\
{\bf PCAw} & {\bf 4}& {\bf no } & $\mathbf{0.79 \pm 0.17}$ &{\bf 0.98} & $\mathbf{4.7\sigma}$\\
mSVD & 4, 8 & no  & $0.91 \pm 0.17$ & 0.99 & 5.6$\sigma$\\
mSVDw & 4, 8 & no  & $0.86 \pm 0.17$ & 1.01 & 5.3$\sigma$\\
{\bf mPCA }& {\bf 4, 8 }& {\bf no}  & $\mathbf{0.93 \pm 0.17}$ & {\bf 0.99} & $\mathbf{5.8\sigma}$\\
mPCAw & 4, 8 & no  & $0.87 \pm 0.17$ & 0.97 & 5.4$\sigma$\\
\hline
{\bf  CL23} & {\bf 30} &{\bf  yes}& $\mathbf{0.86 \pm 0.16}$&{\bf 0.84}& $\mathbf{4.8\sigma}$\\
\hline
\end{tabular}
\tablefoot{Each row corresponds to a different cleaning strategy.  
Being limited by
signal loss, the highest amplitudes are the closest-to-truth, although all amplitude values agree within $\lesssim2\sigma$.
In bold, we highlight: PCA and PCAw to bracket the PCA-like results, mPCA as representative of the multiscale solutions and our benchmark from CL23 (last row), which corresponds to a weighted PCAw with different pre-processing choices and a transfer function correction applied.}
\end{table}

\subsection{Scale-independence of the fitted spectra}

\begin{figure*}
\centering
    \includegraphics[width=1.8\columnwidth]{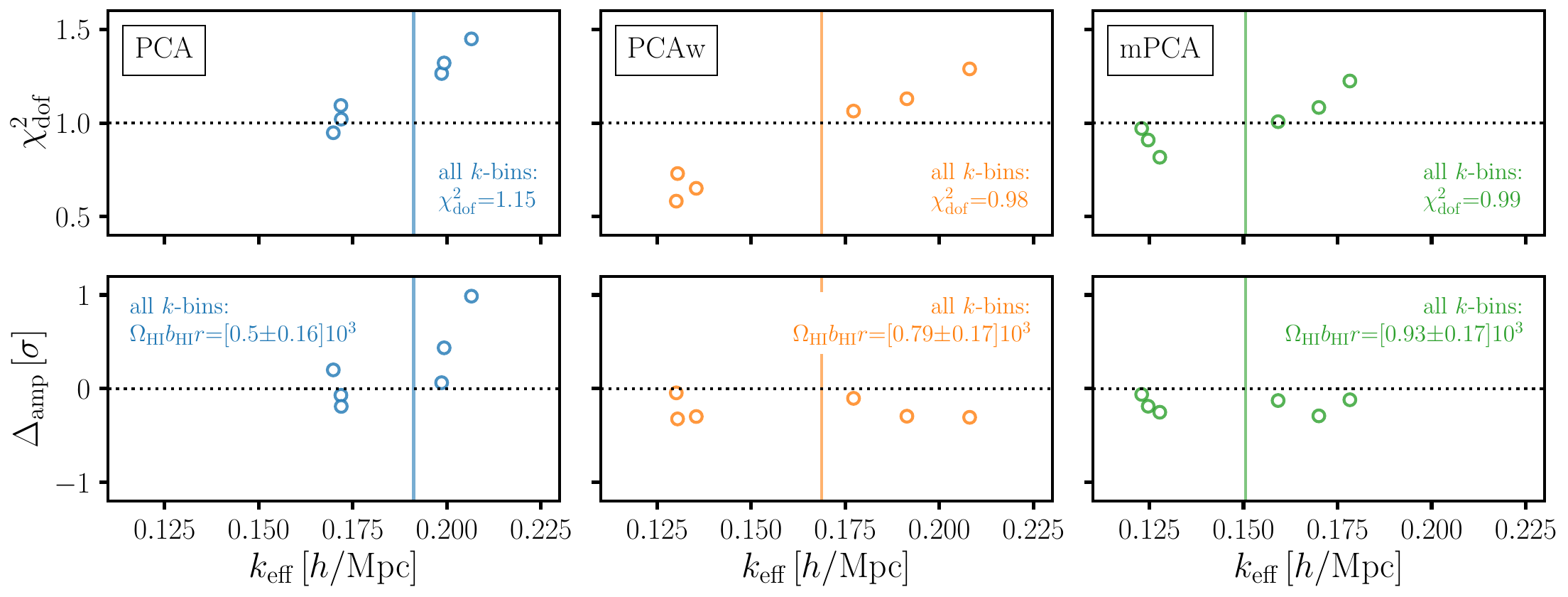}
    \caption{Robustness of the cross-spectra fits. From left to right, we test the scale-independence of the fits for the measured cross-$P(k)$ between the WiggleZ galaxies and the PCA $N_{\rm fg}=4$ cleaned cube (left column), PCAw $N_{\rm fg}=4$ (middle) and mPCA $N_{\rm L}, N_{\rm S}=4, 8$ (right). {\it Top panels:}  Reduced $\chi^2$ for the least-squares analysis as function of the effective scale of the data cube $k_{\rm eff}$ when excluding from 1 to 3 smallest and highest $k$-bins from the fit, namely, by removing either large or small-scale information. {\it Bottom panels:}  Agreement of the amplitudes of the less-information fits with respect to the reference all-$k$-bins fitted amplitude, through the quantity $\Delta_{\rm amp}$, that quantifies how many $\sigma$ away the $\Omega_{\hi} b_{\hi} r$ values are from the reference, indicated in the text for each panel.
    In all panels, the vertical line shows the corresponding  $k_{\rm eff}$ for the reference all-$k$-bin analysis for each cleaning method. 
    } 
    \label{fig:Om_keff}
\end{figure*}

This paper focuses on reaching a compromise between obtaining a good contaminant subtraction while minimising cosmological signal loss. Hence, a test of the robustness of our cross-$P(k)$ measurements is to check their independence with respect to scale; namely, we want to check whether the choice of relying on either large-scale or small-scale information results in significant changes. Indeed, if a certain $k$-scale is more prone to signal loss (typically the large scale), the least-square fit might lead to different solutions when used on a subset of the $P_{\hi,\text{g}}(k)$ sampled $k$-bins.

We tested the scale-independence of the reference cross-spectra obtained with PCA, PCAw, and mPCA,with the results shown in the three columns of Fig.~\ref{fig:Om_keff} (from left to right). For each spectrum, we perform again the least-square fit discarding 1, 2 or 3 $k$-bins either at the low or high end of the available $k$-range; namely, we produce six new fits for each measured $P_{\hi,\text{g}}(k)$, neglecting the large or small scale. For each of these new fits, we can associate an effective scale 
\begin{equation}
    k_{\rm eff} = \frac{\sum_i^{N_k}{k_i \big(P(k_i)/\sigma(k_i)\big)}}{\sum_i^{N_k}{ \big(P(k_i)/ \sigma(k_i)\big)}}\,.
\end{equation}
When discarding low (high) $k$-bins $k_i$, $k_{\rm eff}$ grows (lowers). 
In all panels of Fig.~\ref{fig:Om_keff}, we highlight with a vertical line the $k_{\rm eff}$ of the reference spectra, using all 11 $k_i$ bins: for each column of Fig.~\ref{fig:Om_keff}, we have three data points on the left, three on the right of that line, for the large-scale and small-scale-only fits respectively. We notice that the PCA results (left panels) are in the right area of the plot: the signal-to-noise $P(k_i)/\sigma(k_i)$ is generally worse for low $k_i$ compared to mPCA (right panels); the PCAw case lies between the two (middle panels). 

In the top panels of Fig.~\ref{fig:Om_keff}, we show the $\chi^2_{\rm dof}$ of the fits as a function of $k_{\rm eff}$. In all three scenarios, the low $k_i$ bins are associated with higher relative uncertainties, hence the large-scale fits tend to display lower $\chi^2_{\rm dof}$ than the small-scale fits. Anyway, these new fits show good $\chi^2_{\rm dof}$: the cross-correlation theoretical model describes data reasonably well regardless of the $k$-scale we analyse. The latter is especially true for the mPCA case (last column).

In the bottom panels of Fig.~\ref{fig:Om_keff}, we compare the fitted amplitudes of $\Omega_{\hi} b_{\hi} r$, that for brevity we call $A^i$, against the reference $A^{\rm ref} \pm \sigma_A^{\rm ref}$, namely, obtained with the all $k$-bins fit reported in the text inside each panel---through the quantity:
\begin{equation}
    \Delta_{\rm amp}^i = \frac{A^i - A^{\rm ref}}{\sigma_A^{\rm ref}}\,.
\end{equation}
With $\Delta_{\rm amp}$, we quantify the agreement of the new fits of the robustness test in units of the uncertainty $\sigma_A^{\rm ref}$. All the $6 \times 3 $ fitted amplitudes are within $1\sigma$ from their reference values and, although not shown, with highly similar detection significances---especially true for PCAw and mPCA: the agreement is remarkable, supporting our detection measurements. By contrast, in CL23, the $\Omega_{\hi} b_{\hi} r$ amplitude was changing by a factor of $\sim2$ (with corresponding $\Delta_{\rm amp}>1\sigma$) in a similar $k$ range (Figure 7 therein), leading to the possibility of the presence of spurious signal in the large scale or a $k$-dependence unexpectedly introduced by the $\cal{T}$-correction. In the analysis of this work, when we remove the large scale, results do not change, and even more impressively, when we mainly rely on the large scale, by excluding the small-scale, we still get the same results.

\subsection{Assessing leakages in the component separation}

Our starting point for subtracting contaminants from the MeerKAT intensity maps is the linear decomposition in Eq. (\ref{eq:master}).
Contaminant separation aims to minimise leakages between the sum terms of Eq. (\ref{eq:master}).

\begin{figure}
\centering
    \includegraphics[width=0.85\columnwidth]{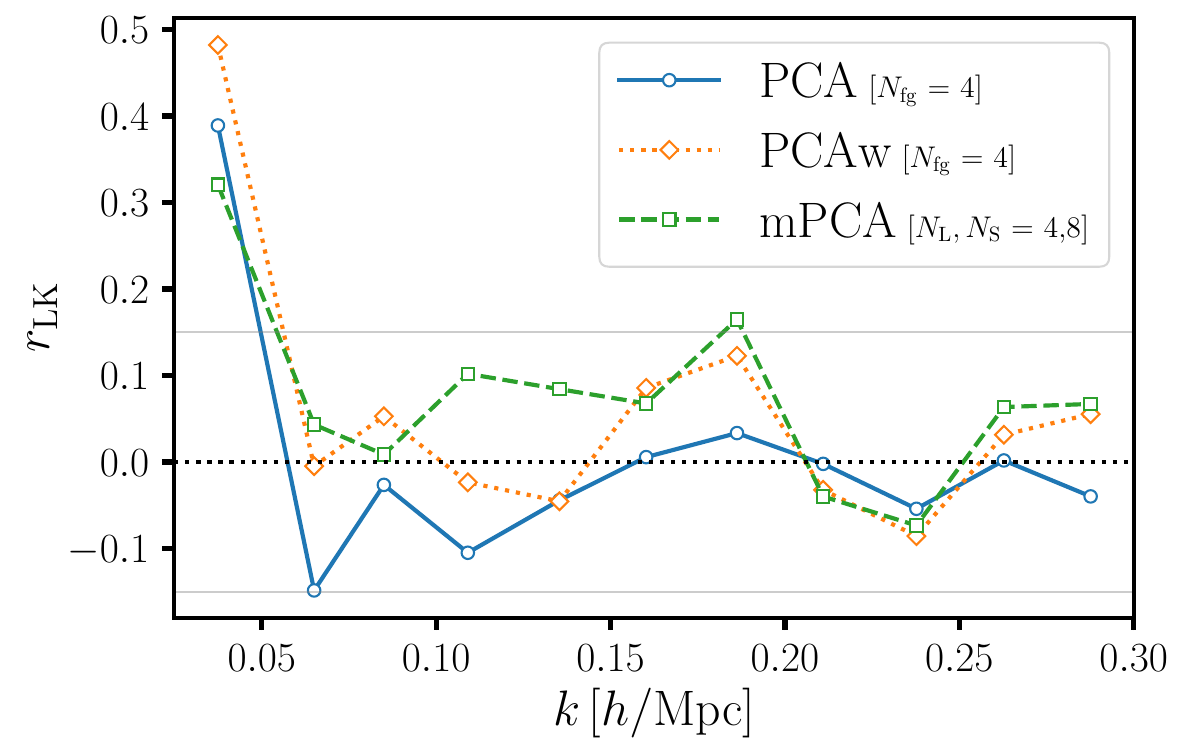}
    \caption{ Cross-correlation coefficient between reconstructed foregrounds and cosmological signal, as defined in Eq. (\ref{eq:rLK}). We show results for the PCA $N_{\rm fg}=4$ cube (solid blue line), PCAw $N_{\rm fg}=4$ (dotted orange) and mPCA $N_{\rm L}, N_{\rm S}=4, 8$ (dashed green). 
    }
    \label{fig:resXFGs}
\end{figure}

For each of the cleaned, reference cubes of this work, we statistically evaluate the leakages between the summed terms of Eq. (\ref{eq:master}) by cross-correlating $\textbf{X}_{\rm clean}$ with the estimated foregrounds $\textbf{X}_{\rm FGs} = \textbf{X} - \textbf{X}_{\rm clean}$. Such a signal could be produced if the contaminants are insufficiently characterised and leak into the residuals or if the removed foregrounds contain some cosmological signal (i.e. signal loss). We compare these cross-spectra with their respective autos through the cross-correlation coefficient, $r_{\rm LK}$:
\begin{equation} \label{eq:rLK}
    r_{\rm LK} = \frac{\mathcal{P}(\textbf{X}_{\rm clean}, \textbf{X}_{\rm FGs})}{\sqrt{\mathcal{P}(\textbf{X}_{\rm clean}, \textbf{X}_{\rm clean}) \mathcal{P}(\textbf{X}_{\rm FGs}, \textbf{X}_{\rm FGs})}}\,,
\end{equation}
where $\mathcal{P}$ denotes the power spectrum computation described in Sect.\,\ref{sec:model} (e.g. following this notation, the cross-spectra earlier measured in this paper would be expressed as $\hat{P}_{\hi,\text{g}} \equiv \mathcal{P}(\textbf{X}_{\rm clean},\textbf{X}_{\rm galaxies})$).
We show the behaviour of $r_{\rm LK}$ for the different cleaned cubes in Fig.~\ref{fig:resXFGs}. 
The correlation between the estimated foregrounds and the estimated cosmological signal oscillates around zero, with a maximum, absolute amplitude of $|r_{\rm LK}| < 0.15$ at $k>0.06\,h/$Mpc for all three (PCA, PCAw, and mPCA) solutions, highlighting a reasonable degree of no-leakage\footnote{The exception is the first $k$-bin sampled that corresponds to $0.3<r_{\rm LK} < 0.5$. However, we recall that when we discard this largest $k$-bin measurement from the final fit, our analysis yields amplitude measurements statistically unchanged (Fig.~\ref{fig:Om_keff}).}.  

\subsection{Assessing signal loss: Mock-injection tests}
\label{sec:sims}

The contaminant separation methods filter out some cosmological signal from the final cleaned maps. This effect is particularly significant on large scales, leading to a suppression of the signal power spectrum. Hence, through mock-injection tests, a corrective transfer function $\cal{T}$ can be constructed. The $\cal{T}$ has been widely used in the analyses of \hi~intensity mapping observations \citep[see e.g.][and \citet{switzer2015} for a more in-depth discussion]{masui2013,Anderson2018,wolz2022,CL23}. In practice, it enhances the power spectrum to counteract the estimated suppression. 

Although we have demonstrated that the cross-power spectrum measurements in this work, given their uncertainties, do not significantly suffer from signal loss because we do not see any evidence of its $k$-dependent effect, we further corroborate that statement by computing the $\cal{T}$ function and correcting the measurements accordingly.

\begin{figure}
\centering
    \includegraphics[width=0.9\columnwidth]{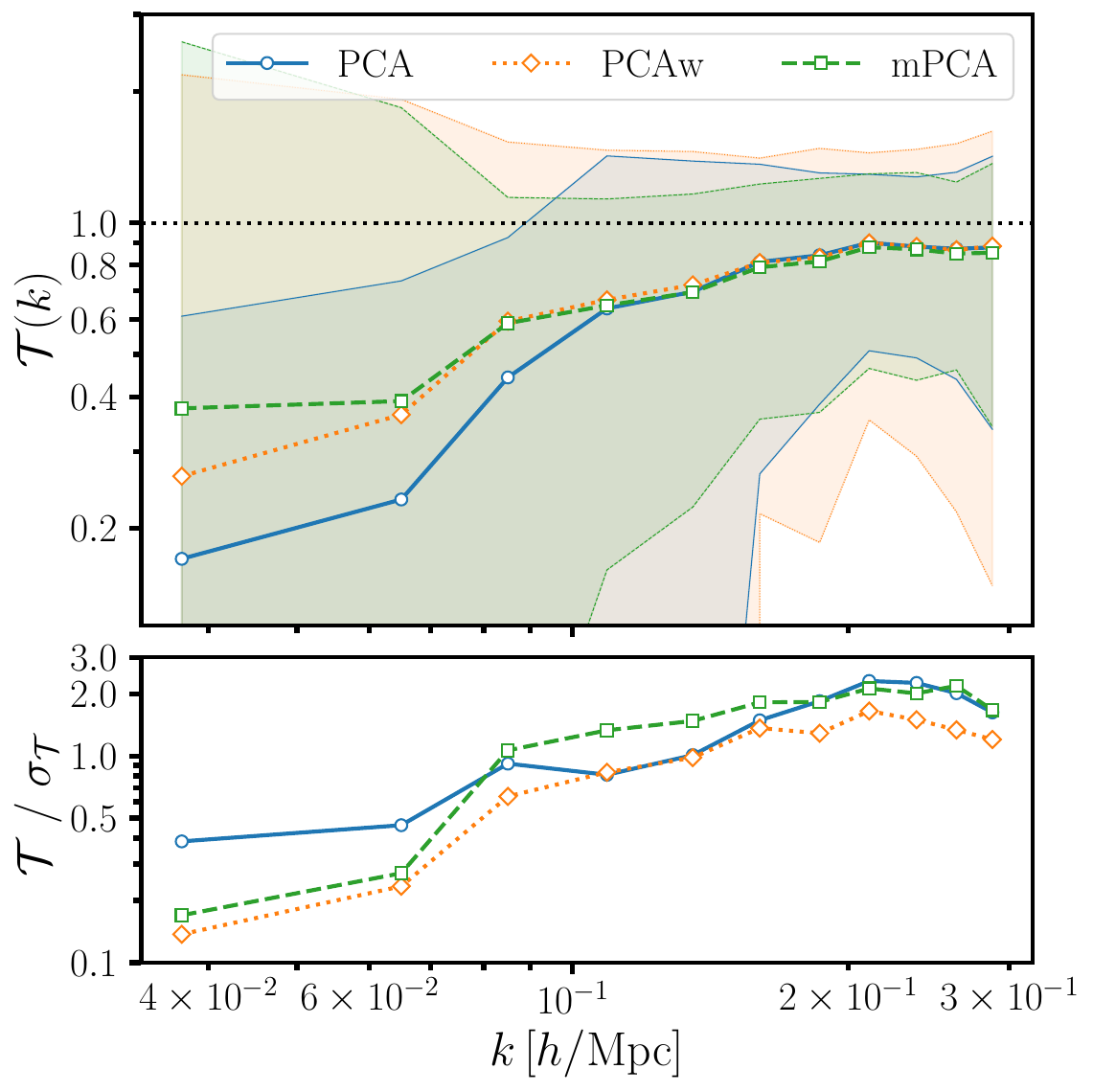}
    \caption{{\it Top panel} Transfer function $\mathcal{T}(k)$ of the three reference analyses: PCA $N_{\rm fg} = 4$ (solid blue), PCAw $N_{\rm fg} = 4$ (dotted orange) and mPCA $N_{\rm L},N_{\rm S} = 4, 8$ (dashed green) and their $1\sigma$ scatter (over 200 mocks) with corresponding shaded areas. {\it Bottom panel:} Signal-to-noise ratio of $\mathcal{T}(k)$, i.e. the ratios over the $1\sigma$ scatter.
    }
    \label{fig:TFtrue}
\end{figure}

\begin{figure*}
\centering
    \includegraphics[width=1.7\columnwidth]{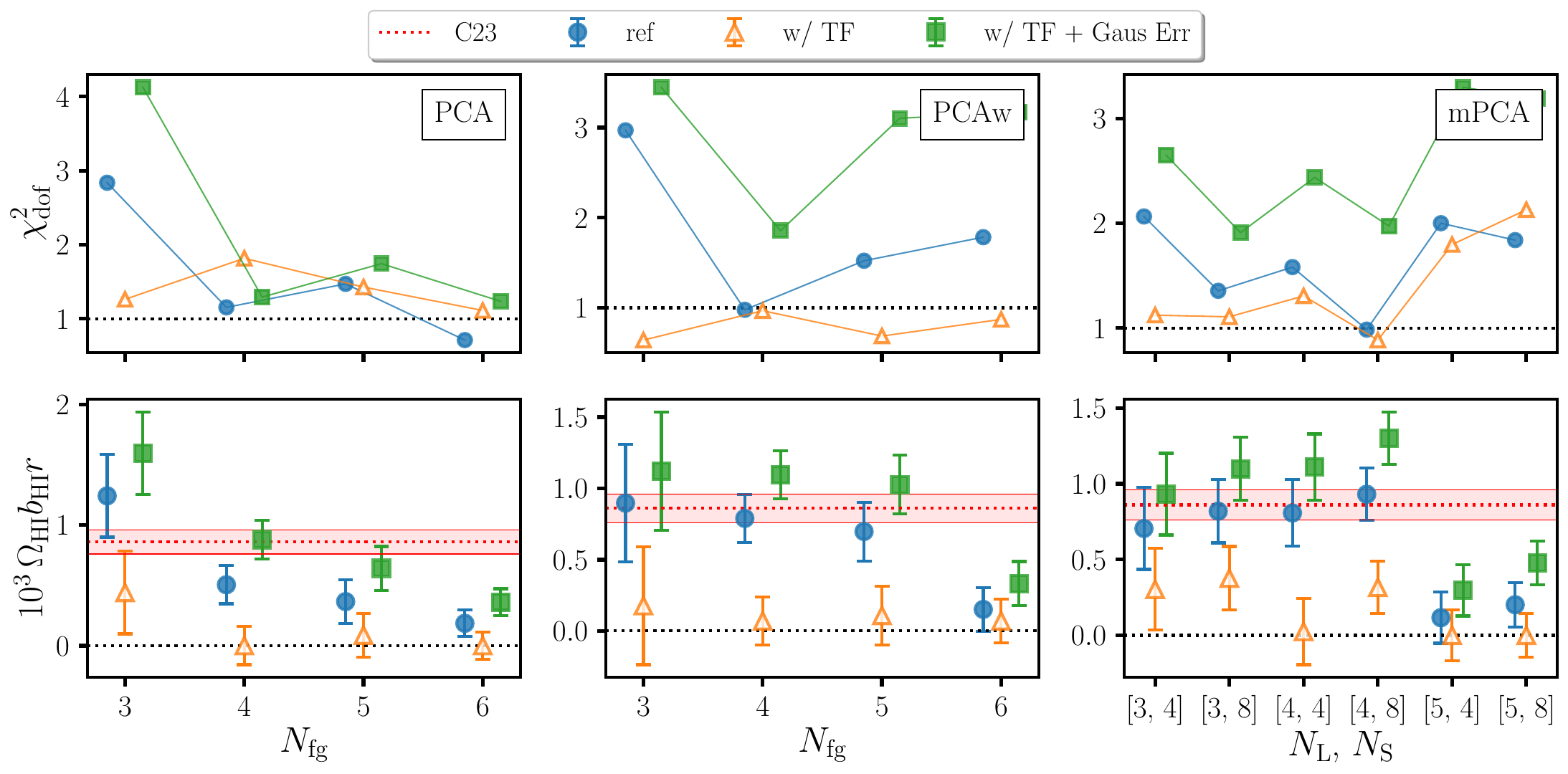}
    \caption{
    Top panels:\ Reduced $\chi^2_{\rm dof}$ of the cross-power best fits. Bottom panels: Corresponding amplitude values compared to the CL23 result in red, both as function of the cleaning level $N_{\rm fg}$. Different columns correspond to different methods: PCA, PCAw and mPCA from left to right. The filled blue circles correspond to the `standard' reference analysis so far discussed; the empty orange triangle to the analysis performed with the transfer function correction; the filled green squares correspond  to the analysis performed with the transfer function correction and imposing Gaussian uncertainties for the final, corrected measurements.}
    \label{fig:TF_fits}
\end{figure*}

We followed the recipe derived by \citet{cunnington2023}, who also validated a method that uses the transfer function variance for error estimation for the power spectra measurements. We began by generating $N_{\rm m}$ mock \hi~signal cubes \textbf{X}$_{\rm m}$, in boxes of same size and gridding as our observed cube; we followed the lognormal approximation \citep{Coles1991} and sampled from the theoretical power spectrum model in Eq. (\ref{eq:Pkmodel}), with cosmology and beam parameters as in our fitting procedure, and \hi~bias and mean brightness temperature values of CL23, assuming $r=1$. Each mock is then injected into the actual data. We run our component separation pipelines (PCA, PCAw, or mPCA) on the combination \textbf{X}+ \textbf{X}$_{\rm m}$. Thus, for mPCA, this process also entails going through the wavelet-filtering and back. In short, for each \textbf{X}+ \textbf{X}$_{\rm m}$ and each method, we determine the mixing matrix $\hat{\textbf{A}}_{\rm m}$ that yields the cleaned cube,
\begin{equation}
   \textbf{X}^{\rm m}_{\rm clean} = (\textbf{X} + \textbf{X}_{\rm m})-\hat{\textbf{A}}_{\rm m}(\hat{\textbf{A}}_{\rm m}^\intercal \hat{\textbf{A}}_{\rm m})^{-1}\hat{\textbf{A}}_{\rm m}^\intercal (\textbf{X} + \textbf{X}_{\rm m}) -[\textbf{X}_{\rm clean}]  \,.
\end{equation}
The last term in the square brackets is the original cleaned observed
data, which we subtract when we want to remove contributions in the cube uncorrelated to the mock signal, thus reducing the variance of the $\cal{T}$ function. In practice, we subtract it when computing the $\mathcal{T}(k)$ but not when computing its covariance.
The transfer function is defined as:
\begin{equation} \label{eq:TF}
    \mathcal{T}(k) =  \left\langle \frac{\mathcal{P}(\textbf{X}^{\rm m}_{\rm clean},\textbf{X}_{\rm m})}{\mathcal{P}(\textbf{X}_{\rm m},\textbf{X}_{\rm m})}  \right\rangle_{N_{\rm m}}\,,
\end{equation}
where $\mathcal{P}$ denotes the power spectrum computation described in Sect.\,\ref{sec:model};  
the external brackets in Eq. (\ref{eq:TF}) correspond to taking the median over $N_{\rm m}$ iterations, one for each mock.

Applying the transfer function correction on the measured $P(k) = \mathcal{P}(\textbf{X}_{\rm clean},\textbf{X}_{\rm galaxies})$ boils down to the following:
\begin{equation} \label{eq:Ptf}
    P_{\rm TF}(k) = \frac{P(k)}{\mathcal{T}(k)} \,,
\end{equation}
with uncertainties of
\begin{equation}\label{eq:Ptf_err}
    \sigma[P_{\rm TF}(k)] = \sigma\{\frac{P(k)}{\mathcal{T}_i(k)} \}\,,
\end{equation}
where $\mathcal{T}_i$ is the transfer function from the $i^{\rm th}$ mock iteration and with $\sigma\{\,\}$ we take the scatter over all iterations.

Results are in Fig.~\ref{fig:TFtrue}, for PCA in solid blue, wPCA in dotted orange and mPCA in dashed green, with the shaded areas corresponding to their $1\sigma$ scatter. The behaviour of the three $\mathcal{T}$ functions is as expected: 
very pronounced---high eventual correction---at the smallest $k$ and quickly increasing to approach 1 at the greatest $k$. 
It is reassuring to see that the pipelines' suppression levels are in the same order as their cross-spectra amplitudes, namely, the lowest measured amplitude value is associated with the highest signal suppression in simulations. Different signal loss levels are likely the reason for the differently cleaned cubes to lead to different cross-spectrum amplitudes.

What is striking in Fig.~\ref{fig:TFtrue} is the amount of variance related to the transfer functions. The plot shows results using  $N_{\rm m} = 200$  mocks; they converge beyond this number. To highlight the $\mathcal{T}$-driven uncertainties, we show the signal-to-noise ratio,  $\mathcal{T}/\sigma_{\mathcal{T}}$, in the bottom panel of Fig.~\ref{fig:TFtrue}, which is below unity for the first half of the $k$-bins we sample and hardly reaches $2-3$ at the smallest scale. Understanding the reason why the $\mathcal{T}$ calculations are so 
undetermined in our analyses is beyond the scope of the present work. Recently, \citet{zhao_quadratic} used a quadratic estimator approach and simulations to demonstrate that, in the presence of mode-mixing, the $\mathcal{T}$ correction is (i) potentially biased and (ii) leads to an incorrect covariance estimation. Here, we have the  chance to test these findings with observations.

We proceed to correct the measured spectra with $\mathcal{T}$ as in Eq. (\ref{eq:Ptf}) and assign errors following Eq. (\ref{eq:Ptf_err}); we re-run the least-square fitting pipeline and determine the best-fitting amplitude values for the different spectra $P_{\rm TF}(k)$. We show results in Fig.~\ref{fig:TF_fits}. As in previous similar figures, we show in the top the goodness of the fits, and in the bottom the fitted amplitudes, both as a function of the number $N_{\rm fg}$ of modes removed. The outcome of the  $P_{\rm TF}(k)$ fits are marked with empty orange triangles, to be compared to the standard analysis (no $\mathcal{T}$-correction) with filled blue circles. Results are equivalent for all methods (PCA, PCAw and mPCA from left to right): uncertainties are so large that all $P_{\rm TF}(k)$ are in agreement with a zero correlation. 

Hypothesising that something is misguided with the uncertainties only (one of the findings of \citet{zhao_quadratic}) and that the $\mathcal{T}$ values themselves are reliable estimates of the clustering suppression, we perform the correction as in Eq. (\ref{eq:Ptf}), but without using the $\mathcal{T}$-variance driven errors; instead, we arbitrarily apply Gaussian statistics to the uncertainties and use Eq. (\ref{eq:Pkerr}) for the $P_{\rm TF}(k)$ measurements, substituting $\hat{P}_{\hi,\text{g}}$ and $\hat{P}_{\hi}$ with the enhanced, $\mathcal{T}$-corrected spectra. Results of this latter exercise are shown as green squares in Fig.~\ref{fig:TF_fits}. The only case for which fits are as reasonable as for the uncorrected spectra is the PCA analysis, in the left column. Interestingly, the new $N_{\rm fg} = 4$ amplitude has increased and is now in better agreement with CL23 and the analyses of other methods.  However, the transfer function correction should work for all PCA analyses with different $N_{\rm fg}$ that result in differently suppressed signal clustering, and all the corrected $P_{\rm TF}(k)$ should be unbiased, that is, pointing to the same  reconstructed amplitude. However, we do not see this for the PCA $N_{\rm fg}$ = 4, 5, and 6 solutions, which continue to decrease as function of $N_{\rm fg}$, like the uncorrected, standard values (blue circles). Finally, for PCAw and mPCA, these last fits' goodness has worsened compared to the references; in terms of spectra amplitudes, results are less than $1\sigma$ away from the reference values, in agreement with the claim that, given our uncertainties, these cubes do not experience important signal loss effects.

In summary, our computed transfer functions are biased and linked to unreliable covariances, supporting the theoretical work by \citet{zhao_quadratic}. We likely still lack a method to compute $\mathcal{T}$ correctly and propagate uncertainties in this data regime; namely, low signal-to-noise ratios and high corruption by systematics that lead to mode-mixing.
However, we do not want to dedicate efforts to strengthening the $\mathcal{T}$  computation since (i) this work is focussed on comparing the cleaning methods and the most optimal pre-processing choices they require 
and (ii)  we are currently forward-modelling the signal loss effect in the upcoming MeerKLASS observations' analyses, finding it a viable method since data quality and signal-to-noise ratios are steadily improving, leaving constraining power for additional nuisance parameters.

\section{Peering towards an independent hydrogen intensity mapping detection with the mPCA cleaning}
\label{sec:mPCAoutlook}

\subsection{Assessing the cubes' variance in retrospect}
\label{sec:back2variance}

\begin{figure}
\centering
    \includegraphics[width=0.9\columnwidth]{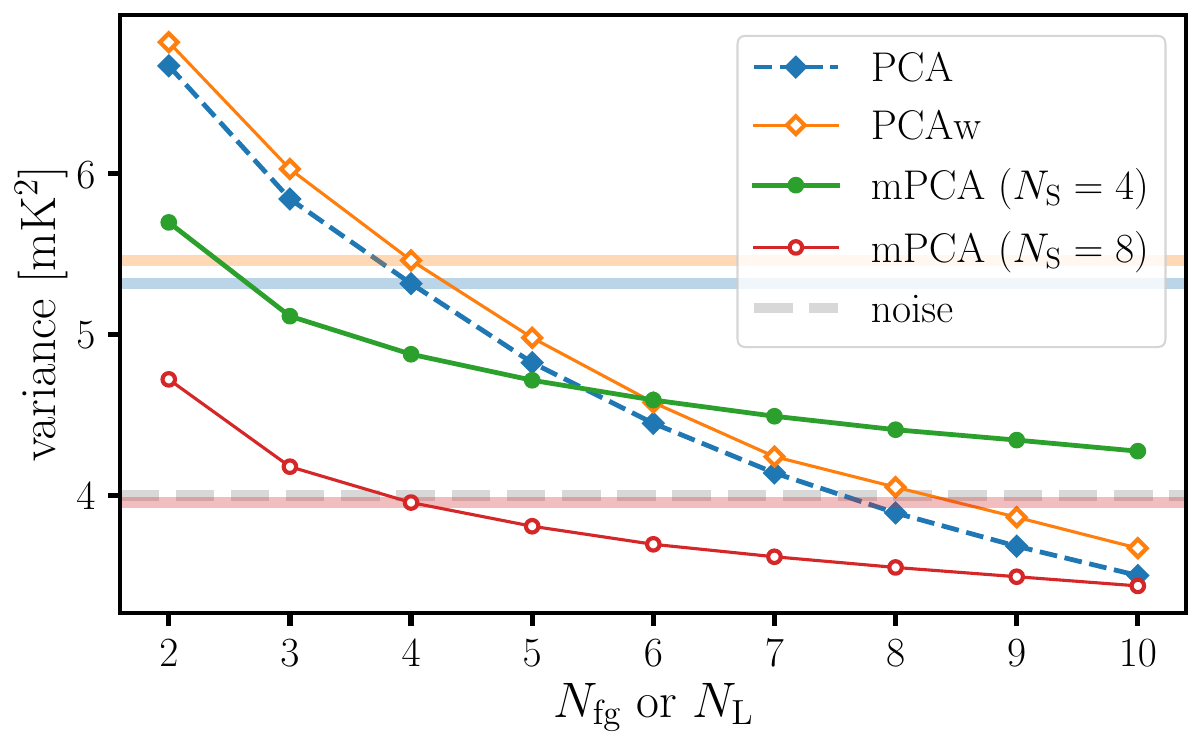}
    \caption{Variance in the cleaned cubes as a function of the number of components removed:\ $N_{\rm fg}$ or $N_{\rm L}$for PCA (dashed blue line), PCAw (solid orange), and mPCA (solid green setting $N_{\rm S}=4$ and solid red for $N_{\rm S}=8$). Horizontal stripes with corresponding colours highlight the variance level of the reference cube for each method ($N_{\rm fg}=4$ for PCA and PCAw and $N_{\rm L},N_{\rm S}=4,8$ for mPCA), namely the solutions for each method that offer a maximised goodness-of-fit and amplitude when cross-correlated with the galaxy catalogue.  For reference, the variance of the original,  unclean cube corresponds to $\approx9 \times 10^3 {\rm mK}^2$ and is characterised by the noise level highlighted by the dashed gray stripe. 
    } 
    \label{fig:var_summary}
\end{figure}

As discussed earlier, in a PCA framework, the quicker (with a lower $N_{\rm fg}$) we can reduce the variance of the data cube, the more efficient the decomposition. If removing an additional component does not visibly change the variance, the assumption of PCA itself is no longer valid, as the principal components should `explain' variance.

In retrospect, after having measured the cross-spectrum with the external tracer, we can compare the best-cleaned cubes' variances. We do so in Fig.~\ref{fig:var_summary}, yielding counter-intuitive results. The cube that retained most of the cosmological signal---higher cross-correlation amplitude---is not that with the overall highest variance, namely, the red, transparent horizontal stripe of mPCA, compared to PCA (blue) and PCAw (orange). Hence, the higher variance of PCA and PCAw is likely linked to higher contaminant residuals in the maps than mPCA. The latter hypothesis is supported by the robustness tests' results shown in Fig.~\ref{fig:Om_keff}.  Foreground residuals will become problematic as we independently measure the \hi~intensity mapping signal: in such cases, we would need more aggressive cleaning, further worsening the signal loss problem.

In Fig.~\ref{fig:var_summary}, we also plot the $N_{\rm S}=4$ mPCA solutions as a function of $N_{\rm L}$, to highlight how, although not fully appreciable in the cross-correlation measurement,  $N_{\rm S}$ plays a crucial role in the mPCA decomposition.

\subsection{Correlation matrices: Off-diagonal terms}

\begin{figure}
\centering
    \includegraphics[width=0.8\columnwidth]{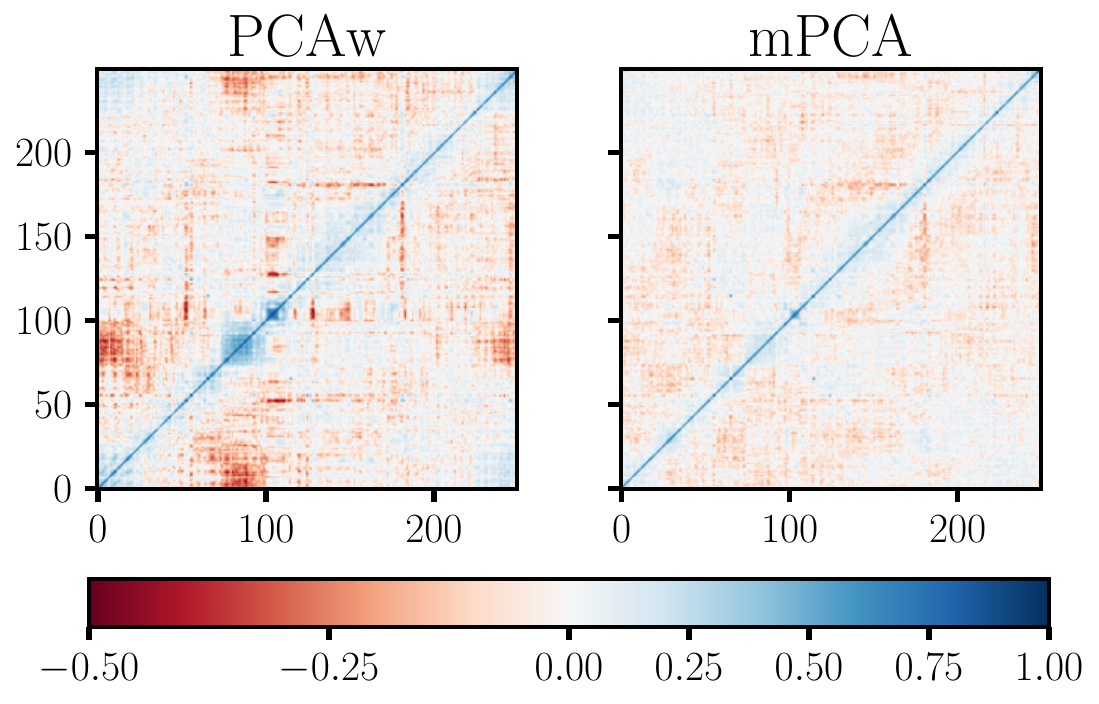}
    \caption{Frequency-frequency correlation matrices of the cleaned cubes, PCAw with $N_{\rm fg}=4$ on the left and mPCA with $N_{\rm L}, N_{\rm S}=4,8$ on the right. Zero correlation corresponds to the white colour. Both matrices show the expected positive diagonal structure of the \hi~IM signal. The mPCA solution results in fewer off-diagonal contributions. 
    } 
    \label{fig:corr_matrix}
\end{figure}

We plot the normalised $\textbf{X}_{\rm  clean}\textbf{X}_{\rm clean}^\intercal$ matrices in Fig.~\ref{fig:corr_matrix}. On the left panel, we show the solution of the PCAw analysis with $N_{\rm fg} =4$; 
on the right panel, the mPCA solution with $N_{\rm L}, N_{\rm S}=4,8$. Both matrices refer to cleaned cubes that show positive and comparable correlations with the WiggleZ catalogue. Yet, they look different: the matrix on the right (mPCA) is substantially `cleaner' than its PCAw counterpart: the diagonal is positive, as expected, and with a much stronger correlation than any off-diagonal term. The secondary tracer---the galaxies---allowed us not to worry about the spurious correlations. We expect to start worrying about them as we proceed toward an independent clustering detection with the \hi~temperature fluctuations maps.

\section{Discussion}
\label{sec:discussion}

\subsection{Confirmation of the detection in cross-correlation}

The current study enhances the CL23 detection of the cross-correlation signal between the MeerKAT \hi~intensity maps and the WiggleZ galaxies at about $z\approx0.4$. We summarise the cross-$P(k)$ detections, significances and fitted amplitudes in Table~\ref{table:summary}.

Although the data are the same as in CL23, pre-processing steps (fewer cuts in frequency space, more in pixel space, no reconvolution) and cleaning method details, especially for the novel multiscale component separation analysis, have changed. Actually, we used the cross-correlation of CL23 as a benchmark to test our signal-processing choices, and the exercise has led to improvements in our analysis pipeline, as we discuss below. Moreover, in the present analysis, all correlation disappears after $N_{\rm fg}\approx8$ components removed, compared with the optimal $N_{\rm fg}=30$ in CL23, a further indication that we have made contaminant subtraction more efficient.

Correlating the MeerKAT maps with the galaxy catalogue allows us to be moderate with the suppression of contaminants in the data. The aim is to find the best compromise between cleaning the maps without removing too much cosmological information. The sense of `compromising' is clear if we look back at Figs.~\ref{fig:crossPK} and~\ref{fig:fit_results} for the standard PCA-like methods and Figs.~\ref{fig:crossPK_mPCA} and~\ref{fig:fit_results_mPCA} for the multiscale case. The $N_{\rm fg} = 4$ and $N_{\rm L}, N_{\rm S}=4, 8$ results are the best compromise to, on the one hand, remove enough contamination to `reveal' a cosmological signal that correlates with the galaxies and, on the other hand, limit the cosmological signal loss. Indeed, for $N_{\rm fg} < 4$, no detection is possible (too much contamination still present); for $N_{\rm fg} > 4$, we have detection (up to $N_{\rm fg} \approx 6$) but with gradually worse fits and less significance: we are experiencing cosmological signal loss. Eventually, we remove too much with $N_{\rm fg} > 6$ and see nothing. 

A welcome consequence of the situation described above is that, in our case,  the optimal choice of $N_{\rm fg}$ is well motivated by the analysis: the chosen $N_{\rm fg}$ is the one that maximises the cross-$P(k)$ amplitude and gives plausible fits with the theoretical model. Indeed, the need to (almost) arbitrarily set $N_{\rm fg}$ has been a drawback of these component separation methods\footnote{There have been propositions in the literature to set $N_{\rm fg}$ automatically \citep[e.g.][]{olivari2016}. However, these schemes typically need a prior for the signal, which is hard to provide for first-of-their-kind measurements, as the large scale \hi~IM signal we aim for, but could eventually become advantageous with the maturity of the field.}. This approach should work also when analysing \hi~data only, looking for a direct detection through internal cross-correlations (e.g. among time-split subsets of data), as the MeerKLASS collaboration plans to do in the future.

We corroborated the robustness of the cross-$P(k)$ detection with the goodness of the fit, the comparison with a zero-detection, null tests, and a demonstrated independence from $k$-scale cuts. Yet, what about the amplitude of the measured cross-$P(k)$? How solid are the \hi~properties we derived through the $\Omega_{\hi}b_{\hi}$ parameters, and why, compared to CL23, we did not correct for signal loss?

\subsection{Beyond the state-of-the-art}

The cross-power spectra of this work have not been boosted with a transfer function, $\cal{T}$, correction, yet they fully agree with CL23, where a strong $\cal{T}$ was in place. 
This agreement is remarkable and has significant consequences. Most importantly, the signal loss in the cleaned cubes of this work is small enough to be negligible within the uncertainties that still characterise our analysis, as explicitly shown in Sect.\,\ref{sec:sims} and confirmed by the robustness test on the $k$-independency of our results (Fig.~\ref{fig:Om_keff}). 

We also seek to discuss why the resulting cross spectra of this work do not need a correction for power loss. Firstly, as pointed out in Sect.\,\ref{sec:theory}, the amount of cosmological signal lost during the component separation process grows with the number $N_{\rm fg}$ of components removed. Hence, keeping $N_{\rm fg}$ as low as possible is the most straightforward and orderly way to limit signal loss. Removing one order of magnitude fewer modes in this work than CL23 translates into preserving more signal. 

Moreover, we expect the eventual correction to be $k$-dependent as we typically lose more signal on the large scale---it is where signal and foregrounds overlap the most, making them difficult to disentangle. This strong $k$-dependency is supported by analyses of different observations \citep[][CL23]{Anderson2018,wolz2022} 
and simulations alike. 
Yet, looking at the least-squares fits in Fig.~\ref{fig:Om_keff} reveals that there is no possibility of a $\cal{T}$-correction to significantly alter the small and large $k$ contributions, given the current measurement uncertainties.

In conclusion, the analysis of this work has further corroborated the \hi~properties that we infer from the MeerKAT L-band intensity maps of the 2019 Pilot Survey (as in Table~\ref{table:summary}). It has also gone beyond that since it has demonstrated that we can measure the cross-correlation signal relying on our novel contaminant suppression pipeline that adequately preserves the cosmological signal, is robust to scale-cuts and points to an unambiguous choice of the level of cleaning ($N_{\rm fg}$ parameter). All the latter features are novel in the \hi~intensity mapping observational context and represent a significant leap beyond the state-of-the-art.

\subsection{Investigating what choices lead to the improvement}

Compared to CL23, (i) we are more cautious in pixel space, reducing the footprint OF to the CF by analysing the components PCA determines when working in the OF and discarding roughly $60\%$ of the pixels. However, here, (ii) we use the whole 250-channel range in the entire band (CL23 removed about one-third of the total frequencies at an additional RFI-flagging stage before the cleaning). To summarise, we ended up working with a comparable amount of data; the main difference is the choice of being more severe with the flagging in pixel rather than frequency space, making our cube `coherent' and facilitating the PCA decomposition. Another critical difference is that (iii) we do not reconvolve the maps with a frequency-dependent Gaussian kernel, effectively not enhancing mode-mixing in the observed cube. Finally, (iv) we introduced a novel multiscale cleaning framework that has exceeded the performance of the other pipelines. Below, we summarise the fundamentals of this work's contaminant subtraction analysis.
\paragraph{Handling outliers in pixel space.} The weighting scheme based on the scan hits map is insufficient for taking care of the pixel outliers (i.e. pixels largely contributing to increasing the temperature variance of each map) that make the PCA decomposition deviate\footnote{In the presence of strong outliers in pixel space, PCA `wastes' components to describe those features, and we have demonstrated that down-weighting these pixels does not seem enough to stop this from happening \citep[similar conclusions are also drawn in, e.g. ][]{wolz2017}. Novel and more efficient weighting schemes might be possible too.}. At present, we find it is more efficient to flag those pixels.  
Indeed, we demonstrated the importance of performing the signal separation task in the smaller CF: the PCA solution substantially changes; a flagging in the pixel space done after the separation process does not have the same effect.
\paragraph{Coherence in the frequency direction.} On the other hand, being too aggressive in the channel flagging is detrimental: having discontinuous information in frequency harms the PCA decomposition, which can describe abnormal variations in the frequency direction variations but struggles to detect the artificial `jumps' in the spectral behaviour of the foregrounds. Moreover, dimensionality reduction techniques as PCA work best in the regime in which components are outnumbered by observations. Hence,  we should try to maximise the number of channels we work with:\ a practical exercise showing this with simulations is in \citet{carucci2020}.
\paragraph{Working with the maps' native angular resolution.} CL23 and other works applied a two-dimensional smoothing to the maps before the cleaning to suppress small-scale systematics. When the smoothing is frequency-dependent, although making all maps share the same resolution, it effectively enhances mode-mixing and eventually degrades the PCA decomposition \citep{spinelli2022}.  We can also smooth all maps with a constant Gaussian kernel that, in this case, would act as a filter, losing cosmological information, too. Sacrificing the small-scale information could be an option for improving the performance of some cleaning methods, although, for instance, \citet{switzer2013} warned against reducing the `angular' 2D information elements present in the data cube to preserve the efficacy of a contaminant separation based on Eq. (\ref{eq:master}). Given the already large MeerKAT beam, we did not opt for this choice. We can still filter large and small-scale information with a wavelet kernel and retain them both and even treat them differently within the multiscale framework (i.e. using, at best, the small-scale information and not discarding it: see next point below).
\paragraph{Addressing spatial scales independently.} The large scale is where the cosmological signal and the foregrounds are mostly entangled and hard to separate. On the other hand, we see artefacts in the intensity maps on small scales, due to a mixture of systematics and astrophysical foregrounds. Independent handling of contaminant subtraction tasks on different spatial scales with mPCA has proven successful and eventually addresses the long-standing problem in the literature of having to increase the aggressiveness of the cleaning to reduce power on the large scale while typically underestimating the power at the small scale.

\subsection{Outlook}

We are actively implementing the above best practices in analysing new MeerKAT \hi~intensity mapping data. 
However, different cleaning methods might work better with other pre-processing choices than those suitable for the PCA framework. For instance, we know that  a generalized morphological component analysis \citep[GMCA,][]{Bobin2007} can deal with missing channels, leading to discontinuous mixing matrices \citep{carucci2020}, so additional channel flagging might be beneficial at the cleaning stage\footnote{Because GMCA focuses on maximising the sparsity of the $\hat{\textbf{S}}$ components and has no prior on the mixing $\hat{\textbf{A}}$, the latter matrix can show discontinuous columns, i.e. `jumps' in the spectral structure of the components, to accommodate the lacking frequency information.}. On the other hand,  Gaussian process regression \citep[GPR,][]{mertens2018} relies mainly on the frequency direction information and its modelling; hence, as in the case of the PCA and SVD (as well as the independent component analysis, ICA), it suffers when channels are flagged \citep{soares2022}. Moreover, different strategies have been proposed within the PCA framework to make them more robust against outliers \citep{Zuo2019} and more flexible, for example by allowing non-linear, functional bases for the decomposition \citep{Irfan2021}. Room for improvement is present when considering the map resolution, beam deconvolution and beam mismodelling in general that we might want to address before the cleaning; for instance, at the map-making stage \citep{McCallum2021} or even simultaneously \citep{carloni2021}.

Regarding mPCA and any cleaning performed on the wavelet-filtered data, the choice of wavelet scale (i.e. how we define the large and small scales) is subject to the morphology of the maps, hence survey specific. According to the resolved spatial scale range between the map's area and the native resolution of the map, one has to test the appropriate choice. For instance, we expect to test and increase the wavelet scales used when extending future survey areas. The choice of the wavelet dictionary, too, is open to discussion \citep{decaro2025}. 

The multiscale approach is generally promising and versatile, and the wavelet filtering delivers an optimal framework for such analysis, namely, the separation of scales facilitates the application of hybrid methods. Here, we choose to keep PCA on both scales analysed, mostly because we want to keep the analysis as simple as possible, while exploring other pre-processing steps; nothing prevents us from mixing different techniques since the two analyses are independent. 
Other works have proposed `hybrid' and multiscale methods too \citep{olivari2016,spinelli2022,podczerwinski2023,Dai2024}, indeed finding them adaptable to the \hi~intensity mapping context. In general, a contaminant separation method is more efficient than another if its assumptions suit the data better. Hybrid approaches have the potential to retain the advantages of each of the methods that compose them.  

Similarly to PCA, many foreground removal techniques applied in the radio intensity mapping context are based on an eigendecomposition of the data matrix and might benefit from the considerations explored in this article. For example, the Karhunen-Lo\`eve (KL) transform is also called the `signal-to-noise eigendecomposition' because it uses priors for the covariances of the cosmological signal and foregrounds (and their ratios) to assign the different eigenvectors to either parts of the observed signal \citep{Bond1995,Shaw2014}. We find it misleading that sometimes, in the literature, those methods are called `linear' methods as opposed to PCA-like methods which are `non-linear': after all, they all try to isolate the subspace of data where the \hi~signal overthrows contaminants, with their definition of the filter being different---in the PCA framework the `filter' is unambiguously derived from data through the mixing matrix, that is applied linearly to data to remove contaminants.  Similarly to the KL decomposition, methods like GPR are also based on the assumption of knowing at least the `shapes' of the different covariances of the signal and contaminant terms. New adaptations provide for more flexibility \citep[e.g.][]{Diao2024}, and 
a new machine-learning version of GPR has further relaxed this assumption by allowing the covariances to be learned from simulations \citep{Mertens2024}. Works like the one presented here will help building better simulations and eventually allow us to use simulation-based methods in general. 

Finally, we stress again that separation methods such as PCA and mPCA are prone to signal loss. 
Keeping the loss as low as possible renders results more reliable regardless of how we address it, so it should always be pursued. For instance, in this work, we reached a low enough loss level to be negligible with our error bars. In perspective, together with pursuing more careful contaminant subtraction, we would need more robust correction approaches or forward-modelling of this effect. For example, we point out that the literature lacks an analysis of the transfer function framework with, for instance, a realistic beam treatment, a crucial aspect that needs attention (some hints of this are in \citet{cunnington2023}). Indeed, \citet{zhao_quadratic} pointed out that mode-mixing (for which the beam is partly responsible) is the pitfall of the transfer function approach as performed so far.  Most importantly, besides the foreground removal process, other pipeline sections (e.g. bandpass calibration and additional filters) can induce signal loss, which must also be estimated, leading to a ‘global’ assessment of the eventual signal loss.
At the same time, as we approach an independent clustering detection
of the \hi~intensity maps, we will need to characterise the contribution
of residual foregrounds in the cleaned data, which will be competing
with signal loss.

\section{Conclusions}
\label{sec:conclusions}

This work aims to optimise the contaminant suppression scheme for the \hi~intensity maps from the MeerKAT radio telescope, building on the cross-correlation detection reported in CL23, which serves as our benchmark. We measured cross-spectra not only in agreement with CL23, but also more robust (scale-independent) and with a higher detection significance. Importantly, we successfully addressed and mitigated longstanding issues associated with the PCA framework analysis, specifically, the signal loss and ambiguity in selecting the cleaning level.  Overall, this work enhances the confidence in the cosmological signal that we will extract from the upcoming MeerKAT \hi~intensity mapping data, particularly with respect to preserving the largest scales we can sample.

We have carefully revisited our contaminant subtraction pipeline based on PCA decomposition, reaching a good level of cleaning with a moderate number of removed modes,  $N_{\rm fg}$, thereby resulting in less signal loss. 
Minor pipeline differences 
in the \hi~intensity mapping context can lead to significant differences, as we have experienced first-hand. In this work, we attempted to describe those differences and what changes they bring to the resulting cleaned maps and our interpretations of why this is so.

We have introduced the multiscale mPCA method and are getting encouraging results. Compared to the other pipelines studied, mPCA: (i) delivered the highest amplitude of the cross-correlation (being limited by signal loss, the highest amplitude is the closest-to-truth, Table~\ref{table:summary}); 
(ii) is associated with less signal loss when looking at simulations (Fig.~\ref{fig:TFtrue}) and to (iii) less prominent off-diagonal terms when analysing the cleaned cube correlation matrix (Fig.~\ref{fig:corr_matrix}) and (iv) lower overall variance (Fig.~\ref{fig:var_summary}), likely linked to less residual contaminants; (v) is demonstrated to be less sensitive to pre-processing choices (mean-centering and weighting), making it overall more robust; finally, (vi) by splitting the contaminant separation problem into two (at large and small scales), mPCA gives more flexibility in the overall separation process (e.g. Fig.~\ref{fig:var_summary}).
In short, the mPCA approach has revealed improved component separation compared to its whole-scale counterpart PCA, with the prospect of independently detecting a cosmological signal using the  intensity map fluctuations. 

The focus of this work has been the subtraction of contaminants; hence, we decided to leave everything else unchanged with respect to analysis described in CL23. However, we expect (and we have experienced first-hand) that the data quality will improve independently of the cleaning because of all the analysis development the collaboration is working on. These works include new calibration schemes and RFI handling, and satellite contamination assessment, as well as new re-gridding schemes and map-making \citep{MK_GAMA,Cunnington2024_grid,Engelbrecht2024}.

PCA is still the workhorse for the \hi~intensity mapping cleaning challenge; it is intuitive, computationally cheap, and mathematically sound. Nevertheless, it cannot blindly be applied to data, as its results rely on specific pre-processing choices. The by-products of the cleaning (i.e. the mixing matrix and the components removed) are excellent indicators of the component separation efficiency and their interpretation can guide the modelling of foregrounds and systematics and the development of more sophisticated pipelines, both `blind' and `forward-modelling' alike.


\begin{acknowledgements}
The authors would like to acknowledge all members of the MeerKLASS collaboration for the valuable insights and feedback from the ongoing analysis work. IPC would also like to thank Jérôme Bobin for stimulating her interest in component separation and statistical learning.

IPC is supported by the European Union within the Next Generation EU programme [PNRR-4-2-1.2 project No. SOE\textunderscore0000136, RadioGaGa] and acknowledges support in the early stages of this work from the `Departments of Excellence 2018-2022' Grant (L. 232/2016) awarded by the Italian Ministry of University and Research (\textsc{mur}).
JLB acknowledges funding from the Ramón y Cajal Grant RYC2021-033191-I, financed by MCIN/AEI/10.13039/501100011033 and by the European Union “NextGenerationEU”/PRTR, as well as the project UC-LIME (PID2022-140670NA-I00), financed by MCIN/AEI/ 10.13039/501100011033/FEDER, UE.
SC is supported by a UK Research and Innovation Future Leaders Fellowship grant [MR/V026437/1]. 
MGS acknowledges support from the South African National Research Foundation (Grant No. 84156).
JW acknowledges support from the National SKA Program of China (No. 2020SKA0110100). 
JF acknowledges support of Funda\c{c}\~{a}o para a Ci\^{e}ncia e a
Tecnologia through the Investigador FCT Contract No.
2020.02633.CEECIND/CP1631/CT0002, and the research grants
UIDB/04434/2020 and UIDP/04434/2020.
YL acknowledges the support of the National Natural Science Foundation of China (grant No. 12473091).
AP is a UK Research and Innovation Future Leaders Fellow [grant MR/X005399/1].

We acknowledge the use of the Ilifu cloud computing facility, through the Inter-University Institute for Data Intensive Astronomy (IDIA).
The MeerKAT telescope is operated by the South African Radio Astronomy Observatory, which is a facility of the National Research Foundation, an agency of the Department of Science and Innovation.
\end{acknowledgements}

%
\bibliographystyle{aa} 
\bibliography{MK_mPCA.bib} 

\begin{appendix}





\section{Testing channel flagging and reconvolution}
\label{app:test}

Some pre-processing choices in the present analysis were justified by the evidence in other---simulation-based---works (Sect.\,\ref{sec:steps}). Here, with the cross-$P(k)$ measurements, we can test them on data and further support those findings.

\begin{figure*}[bp]
\centering
\includegraphics[width=1.7\columnwidth]{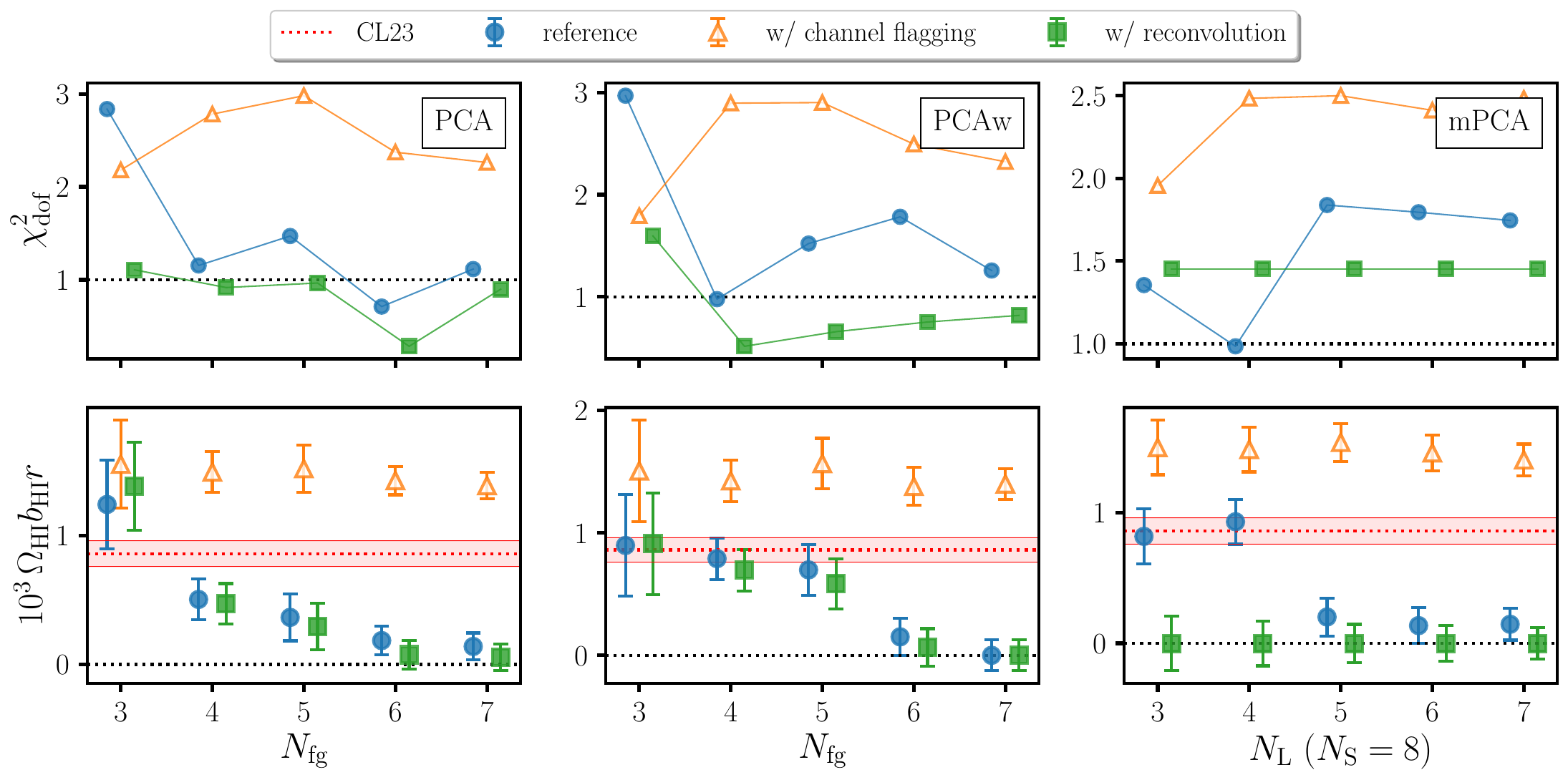}
    \caption{
    Top panels:\ Reduced $\chi^2_{\rm dof}$ of the cross-power best-fits. Bottom panels:\ Corresponding amplitude values compared to the CL23 result in red, as functions of the cleaning level. 
    Columns correspond to PCA, PCAw and mPCA from left to right. The filled blue circles correspond to the `standard' reference analysis; the empty orange triangle to the analysis  with the extra flagging of $\sim 20\%$ of the channels 
    before the component separation; the filled green squares correspond to the analysis performed reconvolving all maps before the component separation.
    } 
    \label{fig:fit_results_options}
\end{figure*}

\subsection{Channel flagging}
The extra channel flagging 
makes the data cube information discontinuous in frequency. When seen in the context of Eq. (\ref{eq:master}), this translates into looking for a discontinuous mixing matrix, which the PCA-based methods have difficulty determining. We put this into practice by identifying the channels from the CF data cube that exhibit the highest absolute deviation in variance, the latter fitted in frequency with an order 2 polynomial.  
We mimic the CL23 analysis and discard $\sim 50$ channels (maps) from the initial 250. 
We collapse the data cube and continue the analysis. We show the final results from the cross-$P(k)$ fits as a function of $N_{\rm fg}$ removed in Fig.~\ref{fig:fit_results_options} with empty triangles to be compared to the `standard' analysis with filled blue circles: results have deteriorated for all methods (PCA in the left panels, PCAw in the middle, and mPCA on the right). 
The top panels refer to the goodness of the fit, which is underperforming for all cases looked at. Frequency-incomplete data compromise PCA-like contaminant separations. \citet{carucci2020} show with simulations that the separation does not improve increasing 
the number of components removed, as we show here for up to $N_{\rm fg}=7$.

\subsection{Reconvolution}
Reconvolution has been customary in analysing hydrogen intensity mapping experiments \citep[][CL23]{masui2013,wolz2017,Anderson2018,wolz2022}. It is performed to reduce overall data variance and, specifically, to (i) equalise the cube resolution, since component separation is more efficient when maps share resolution, and (ii) smooth small-scale artefacts. However, (i) a proper beam-deconvolution needs an accurate beam model, but typically, reconvolution assumes a Gaussian kernel; (ii) and it potentially increases mode-mixing by altering maps in a frequency-dependent way, hence detrimental to cleaning efficiency. We test it against our measurements here.

We follow CL23. We convolve each temperature map $T(p,\nu)$ at frequency $\nu$ with the following kernel:
\begin{equation}\label{eq:ResmoothKernel}
    \mathcal{K} (\Delta p,\nu)=\exp \left[-\frac{\Delta p^{2}}{2[\gamma \sigma_{\max }^{2}-\sigma_{\rm beam}^{2}(\nu)]}\right]\,,
\end{equation}
with $\Delta p$ the angular separation between pixels, and
$\sigma_\text{beam}(\nu)$ the frequency-dependent resolution when approximating the beam with a Gaussian function as defined in Eq. (\ref{eq:sigbeamsize}).
$\sigma_\text{max}$ is the maximum $\sigma_{\rm beam}(\nu)$ value (corresponding to the lowest frequency in the cube); we set $\gamma=1.2$ as in CL23,  
yielding a frequency-independent $\gamma\theta_\text{FWHM}(\nu_\text{min})\,{=}\,1.82\,\text{deg}$  effective resolution, compared to the native median of 1.28 deg we started with. We normalise the kernel in Eq. (\ref{eq:ResmoothKernel}) so that the sum over pixels is equal to one. We then perform a weighted reconvolution:
\begin{equation}
     T^{\rm reconv}(p,\nu) = \frac{\left[ T(p,\nu)\,w_\hi(p,\nu)\right] * \mathcal{K}(\Delta p,\nu)}{w_\hi( p,\nu) * \mathcal{K}(\Delta p,\nu)}\,.
\end{equation}
Finally, the weight maps too are reconvolved according to
\begin{equation}
    w^{\rm reconv}_\hi(p,\nu) = \frac{\left[w_\hi(p,\nu) * \mathcal{K}(\Delta p,\nu)\right]^2}{w_\hi(p,\nu) * \mathcal{K}^2(\Delta p,\nu)}\,.
\end{equation}
After reconvolution, we continued the analysis, going through contaminant separation and measuring the cross-$P(k)$ with the galaxies. Results are displayed as filled green squares in Fig.~\ref{fig:fit_results_options}. Let us start discussing results for the multiscale mPCA analysis in the last column of Fig.~\ref{fig:fit_results_options}. The reconvolution procedure broke down the scales' separation with the starlet decomposition, the mPCA cleaning failed.

For PCA and PCAw, the analyses performed after reconvolution are still competitive, although the uncertainties on the measurements increased, $\chi^2_{\rm dof}$ are smaller, and the overall amplitudes have also decreased.
The frequency-dependent smoothing does not benefit PCA-like contaminant separation.

\end{appendix}

\end{document}